\documentclass[prd,aps,twocolumn,showpacs,preprintnumbers,amsmath,amssymb,superscriptaddress,nofootinbib,english]{revtex4-1}
\usepackage{graphicx}
\usepackage{epsf}
\usepackage{amsmath}
\usepackage{amssymb}
\usepackage{amsfonts}
\usepackage{times}
\usepackage{epstopdf}
\usepackage[usenames]{color}
\usepackage[normalem]{ulem}
\usepackage[T1]{fontenc}
\usepackage[dvipsnames]{xcolor}

\newcommand{\tcr}{\textcolor{black}}

\newcommand{\tcg}{\textcolor{red}}

\usepackage{graphicx}
\usepackage{dcolumn}
\usepackage{bm}
\usepackage{color}

\begin{document}

\title{Revisiting the screening mechanism in $f(R)$ gravity}

\author{Jian-hua He}
\email[Email address: ]{jianhua.he@brera.inaf.it}
\affiliation{INAF-Observatorio Astronomico, di Brera, Via Emilio Bianchi, 46, I-23807, Merate (LC), Italy}

\author{Baojiu Li}
\affiliation{Institute for Computational Cosmology, Department of Physics, Durham University, Durham DH1 3LE, UK}

\author{Adam~J.~Hawken}
\affiliation{INAF-Observatorio Astronomico, di Brera, Via Emilio Bianchi, 46, I-23807, Merate (LC), Italy}

\author{Benjamin~R.~Granett}
\affiliation{INAF-Observatorio Astronomico, di Brera, Via Emilio Bianchi, 46, I-23807, Merate (LC), Italy}

\pacs{98.80.-k,04.50.Kd}

\begin{abstract}
We reexamine the screening mechanism in $f(R)$ gravity using N-body simulations. \tcr{By} explicitly examin\tcr{ing} the relation between the extra scalar field $\delta f_R$ and the gravitational potential $\phi$ in the perturbed Universe\tcr{, w}e find that the relation between these two fields plays an important role in understanding the screening mechanism. We show that the screening mechanism in $f(R)$ gravity depends mainly on the depth of the potential well\tcr{, and} find \tcr{a useful} condition for identifying unscreened haloes in simulations. We also discuss the potential application of our results to real galaxy surveys.
\end{abstract}

\maketitle

\section{Introduction}

\tcr{Compelling} cosmological observations \cite{1,WMAP,BAOm} show that our Universe is undergoing a phase of accelerated expansion. The leading explanation \tcr{to} this acceleration is a cosmological constant in the framework of General Relativity (GR). Despite its notable success in explaining current cosmological data sets, this standard paradigm suffers from several serious problems: the measured value of the cosmological constant is far \tcr{smaller} than the prediction of quantum field theory, and there is also a coincidence problem as to why the energy densities of matter and the cosmological constant are of the same order today (see, e.g., Ref.~\cite{sean} for a review).

There are suggestions that GR might not be accurate on cosmological scales, and modified gravity theories are proposed as alternatives to explain the cosmic acceleration.  One of the simplest attempts is the so-called $f(R)$ gravity, in which the Ricci curvature $R$ in the Einstein-Hilbert action of general relativity is replaced by an arbitrary function of $R$~\cite{fr1,fr2,fr3,fr4,fr5,fr6,fr7,fr8,fr9,fr10,fr11,fr12}.  $f(R)$ gravity introduces a new scalar field degree of freedom that has profound impacts on cosmology. At the background level, the self interaction of this scalar field allows the theory to produce any cosmic expansion history with \tcr{desired} effective dark energy equation of state $w(a)$. At the perturbed level, the local scalar curvature $R$ does not necessarily follow the matter density field and \tcr{thus} high density \tcr{might} not imply high curvature in $f(R)$ cosmology. If the curvature is  significantly lower than the corresponding GR result for the same density field, the local spacetime will be altered and the model may fail to pass the local tests of gravity. Therefore, for viable $f(R)$ models the standard local space-time should be recovered in high-density regions. \tcr{To this end, a} screening mechanism \cite{Khoury} is  essential and plays an important role in the viability of $f(R)$ gravity.

The aim of this paper is to further investigate this important issue. Instead of studying the screening mechanism based on individual isolated galactic haloes \cite{HuS,Fabian,Lucas}, we will examine the relation between the scalar field, $\delta f_R$, and the gravitational potential, $\phi$, in $f(R)$ cosmolog\tcr{ies, using} N-body simulations. We will demonstrate that this relation plays an important role in understanding the screening mechanism in $f(R)$ gravity. In order to strengthen our argument, we shall study two different $f(R)$ models: \tcr{one} which exactly reproduces the $\Lambda$CDM background expansion \cite{frmodel} and the other being the Hu-Sawicki model (H-S hereafter) \cite{HuS}.

This paper is organized as follows: In Sec.~\ref{model}, we will introduce the details of the $f(R)$ models investigated in this work. In Sec.~\ref{nonlinear}, we will briefly review the technique details of N-body simulations. In Sec.~\ref{special}, we will discuss the distribution of the scalar curvature $R$ in the void regions and the screening mechanism in the high-density regions. In Sec.~\ref{haloes}, we will discuss the screening mechanism in the dark haloes.
In Sec.~\ref{con}, we will summarize and conclude this work.

\section{f(R) model\label{model}}

We work with the 4-dimensional \tcr{modified Einstein-Hilbert} action
\begin{equation}
S=\frac{1}{2\kappa^2}\int d^4x\sqrt{-g}[R+f(R)]+\int
d^4x\mathcal{L}^{(m)}\label{action}\quad,
\end{equation}
where $\kappa^2=8\pi G$ with $G$ being Newton's constant, $g$ is the determinant of the metric $g_{\mu\nu}$, $\mathcal{L}^{(m)}$ is the Lagrangian density for matter and $f(R)$ is an arbitrary function of the Ricci scalar curvature $R$ \cite{fr1,fr2,fr3,fr4,fr5,fr6,fr7,fr8,fr9,fr10,fr11,fr12} (see Refs.~\cite{frreview,review_Tsujikawa} for reviews). It is well known that the functional form $f(R)$ completely specifies the quantitative behavior of a model, in particular how efficient the screening mechanism is. As a result, to better illustrate our points, in this work we will study two different $f(R)$ models as described below.

\tcr{The first model to be considered is proposed by one of us, which} can exactly reproduce the $\Lambda$CDM background expansion history~\cite{frmodel}. \tcr{We call this `our model', and it} is specified by
\begin{equation}
\begin{split}
f(R)&=-6\Omega_d^0H_0^2-\frac{3D\Omega_m^0H_0^2}{p_+-1}\left (\frac{3\Omega_m^0H_0^2}{R-12\Omega_d^0H_0^2}\right )^{p_+-1}\\
&\times{_2F_1}\left[q_+,p_+-1;r_+;-\frac{3\Omega_d^0H_0^2}{R-12\Omega_d^0H_0^2}\right
]\label{fr}\quad.
\end{split}
\end{equation}
The indices in the above expression are given by
\begin{eqnarray}
q_+=\frac{1+\sqrt{73}}{12}\nonumber,\quad r_+=1+\frac{\sqrt{73}}{6}\nonumber, \quad p_+=\frac{5+\sqrt{73}}{12}\nonumber.
\end{eqnarray}
${_2F_1}\left[a,b;c;z\right]$ is the hypergeometric function.
When $c>b>0$, the hypergeometric function has the integral representation
\begin{equation}
{_2F_1}[a,b;c;z]=\frac{\Gamma(c)}{\Gamma(b)\Gamma(c-b)}\int_0^{1}t^{b-1}(1-t)^{c-b-1}(1-zt)^{-a}dt,\label{defhypergeometric}
\end{equation}
where $\Gamma(x)$ is the Euler Gamma function; ${_2F_1}[a,b;c;z]$ is a real function that is well defined in the range $-\infty<z<1$ in this case. $H_0$ is the Hubble constant today. $\Omega_m^0$ is the matter density today and $\Omega_d^0=1-\Omega_m^0$. $D$ is an additional parameter that characterises the $f(R)$ model. For the instability issue as discussed in Ref.~\cite{Ignacy}, $D$ must be constrained as $D<0$. Our model predicts a lower bound for the scalar curvature $R$ \tcr{across} the Universe
\begin{equation}
R\in(4\Lambda,+\infty),
\end{equation}
where
\begin{equation}
\Lambda=3\Omega_d^0H_0^2.
\end{equation}

The other model we consider is the one proposed by Hu \& Sawicki \cite{HuS}, for which
\begin{equation}\label{eq:HSfr}
f(R)=-\Omega_m^0H_0^2\frac{c_1\left(\frac{R}{\Omega_m^0H_0^2}\right)^n}{c_2\left(\frac{R}{\Omega_m^0H_0^2}\right)^n+1}.
\end{equation}
This model is designed to explain the late-time cosmic acceleration without a cosmological constant. In the high-curvature regime, \tcr{where}
\begin{equation}
\frac{R}{\Omega_m^0H_0^2} \gg 1,
\end{equation}
\tcr{however,} $f(R)$ \tcr{actually does} reduce to a phenomenological cosmological constant $2\frac{c_1}{c_2}\Omega_m^0H_0^2\sim4\Lambda$~\cite{HuS}. In the opposite limit, it satisfies $f(R=0)=0$. If one chooses $|f_{R0}|\ll 1$ (where $f_R\equiv df(R)/dR$ and a subscript `0' is used to denote its present-day value), the background expansion of the H-S model is practically indistinguishable from the $\Lambda$CDM model. For simplicity, we will take $n=1$ for the H-S model throughout this work.

\section{N-body simulations\label{nonlinear}}

In this section, we shall briefly summarize the basic equations to be used in $f(R)$ cosmological simulations, as well as the techni\tcr{cal} details of our simulations.

\subsection{Non-linear perturbation equations}

The large-scale structure formation in $f(R)$ gravity is governed by the modified Poisson equation
\begin{equation}
\nabla^2\phi=\frac{16\pi G}{3}\delta \rho-\frac{\delta R}{6},\label{poissonfr}
\end{equation}
and the equation of motion for the scalar field $f_{R}$. \tcr{If} $|f_{R}|\ll1$, its equation approximately becomes
\begin{equation}
\nabla^2\delta f_R=\frac{1}{3c^2}[\delta R - 8\pi G\delta \rho],\label{frpoisson}
\end{equation}
where $\phi$ denotes the gravitational potential, $\delta f_R\equiv f_R(R)-f_R(\bar{R})$, $\delta R\equiv R-\bar{R}$, and $\delta \rho\equiv\rho-\bar{\rho}$. The overbar denotes the background quantities, and $\nabla$ is the derivative with respect to the physical coordinates. Eqns~(\ref{poissonfr}) and~(\ref{frpoisson}) are derived in linear perturbation theory under the quasi-static approximation, but can also be used in the non-linear regime, as long as the fully non-linear relation between $f(R)$ and $R$ is used.

\begin{figure}
\includegraphics[width=3.5in,height=2.8in]{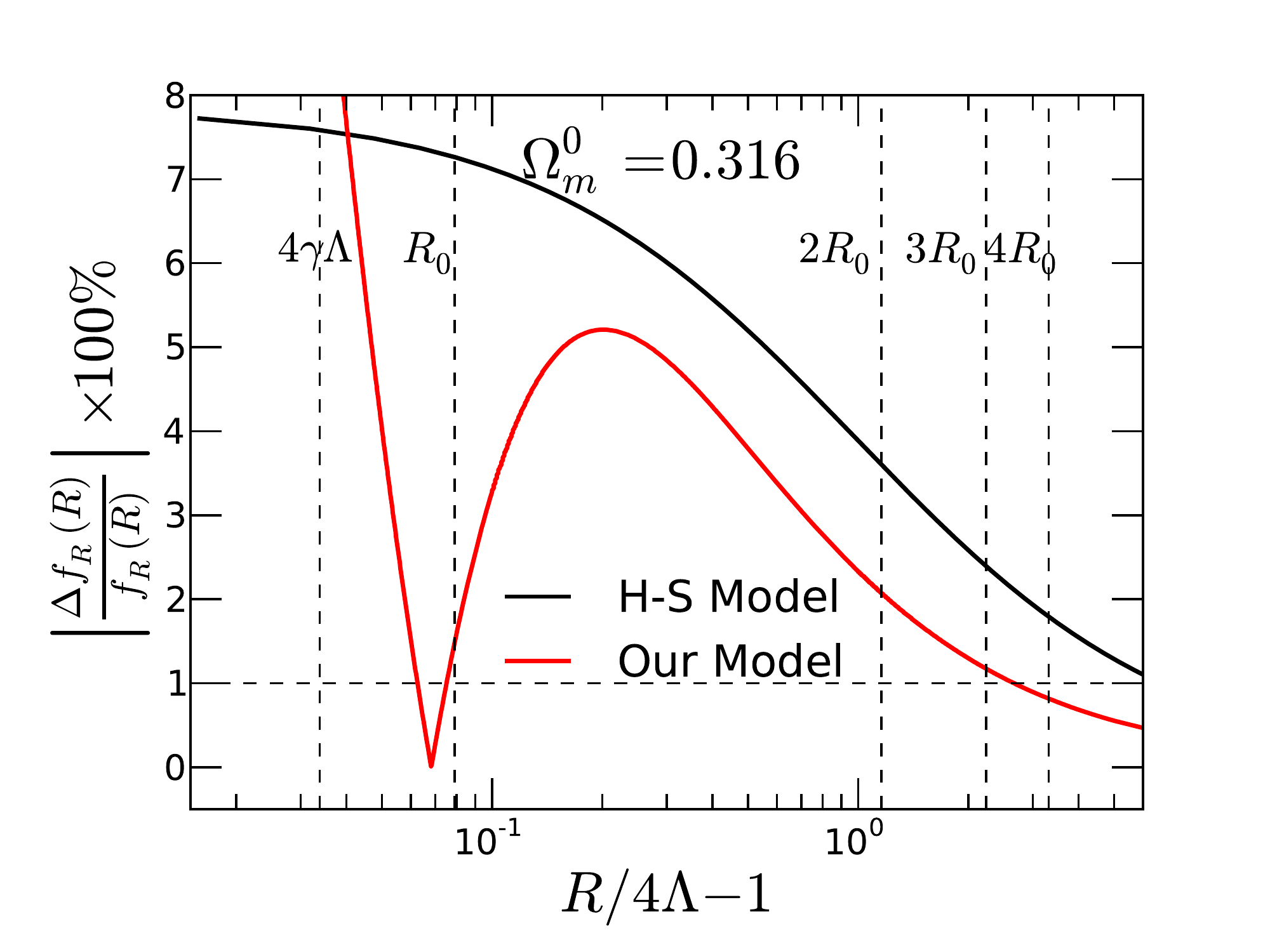}
\caption{The error of the approximation \tcr{for $f_R(R)$} relative to the exact \tcr{expressions}. \tcr{When the curvature is high}, the error in our model drops very quickly. When $R> 3.3 R_0$, the error is below $1\%$. \tcr{When the curvature is low, e.g.}, $R\sim R_0$, the error is around $1.5\%$. The error goes up to $10\%$ when $R$ is around $4\gamma\Lambda$, where $\gamma=1.0338$. However, $4\gamma\Lambda$ is the minimal value of $R$ in our simulations, which actually is a rare case. \tcr{The results show} that the overall accuracy of the \tcr{approximate expression of $f_R(R)$} for our model is better than that for the Hu-Sawicki model with $n=1$. }\label{rel:lower}
\end{figure}

In order to incorporate nonlinear effects into $f(R)$ simulations, we simply need to express $R$ in terms of $f_R$. In practice, however, it is
difficult to do this by inverting the exact expression, Eq.~(\ref{fr}), for our model. Instead, we use a fitting formula
\begin{equation}\label{eq:fitting}
f(R)\sim-6\Omega_d^0H_0^2-\frac{3D\Omega_m^0H_0^2}{p_+-1}\left(\frac{3\Omega_m^0H_0^2}{R-12\alpha\Omega_d^0H_0^2}\right )^{p_+-1},
\end{equation}
where $\alpha$ is a fitting parameter depending on $\Omega_m^0$. Taking the derivative of the above equation, we obtain
\begin{equation}
f_{R}(R)\sim D\left(\frac{3\Omega_m^0H_0^2}{R-12\alpha\Omega_d^0H_0^2}\right )^{p_+}\label{fR:approx}.
\end{equation}
By fitting $\alpha$, Eq.~(\ref{fR:approx}) is found to be relatively a good approximation to the exact derivative of Eq.~(\ref{fr}). In Fig.~\ref{rel:lower}, we show the relative error of our fitting formula with respect to the exact expression, where
\begin{equation}
\left |\frac{\Delta f_R}{f_R}\right| = \left |\frac{f_{R,\rm {app}}-f_{R, \rm {exact}}}{f_{R, \rm{exact}}}\right|.
\end{equation}
In Fig.~\ref{rel:lower}, we find $\alpha=0.9436$ for $\Omega_m^0=0.316$. The relative error between Eq.~(\ref{fR:approx}) and the exact derivative of Eq.~(\ref{fr}) is less than $5.5\%$ for $R>R_0$ where $R_0$ is the Ricci curvature today.  When $R>3.3R_0$, the error drops rapidly down to $1\%$. \tcr{At} $R\sim R_0$, the error is around $1.5\%$\tcr{, and it} only goes up to $10\%$ when $R$ approaches $4\gamma\Lambda$ where $\gamma=1.0338$.  However, as we shall show later, $4\gamma\Lambda$ is the minimal value of $R$ that can be found in our simulations, which is actually very rare.

Using this fitting formula, we can express $R$ in terms of $f_R$ as
\begin{equation}
R=12\alpha\Omega_d^0H_0^2+3\Omega_m^0H_0^2\left(\frac{D}{f_R}\right)^{\frac{1}{p_+}},\label{Rcur}
\end{equation}
\tcr{for our model. As f}or the H-S $f(R)$ model, for $R\gg H_0^2\Omega_m^0$, the scalar field $f_R$ can be approximated as
\begin{equation}
f_R(R)\approx-n\frac{c_1}{c_2^2}\left(\frac{\Omega_m^0H_0^2}{R}\right)^{n+1}.\label{appHI}
\end{equation}
Fig.~\ref{rel:lower} \tcr{also} shows the accuracy of this approximation, and we can see that it is less accurate when $R\sim R_0$, where the error goes up to $7\%$. \tcr{We can similarly invert this equation to get $R$ as a function of $f_R$ for the H-S model, and the final result can be found in, e.g., Ref.~\cite{HuS}.}

\subsection{$N$-body equations}

Our $f(R)$ simulations are performed using the {\sc ecosmog} code \cite{ECOSMOG}, which is itself based on the adaptive mesh refinement (AMR) $N$-body code {\sc ramses} \cite{RAMSES}. The code uses the supercomoving coordinates
\begin{equation}
\begin{split}
\tilde{x}=\frac{x}{aB},\quad\rho=\frac{\rho a^3}{\rho_c\Omega_m^0},\quad \tilde{v}=\frac{av}{BH_0},\nonumber\\
\tilde{\phi}=\frac{a^2\phi}{(BH_0)^2},\quad d\tilde{t}=H_0\frac{dt}{a^2},\quad \tilde{c}=\frac{c}{BH_0},
\end{split}
\end{equation}
where $x$ is the comoving coordinate, $\rho_c$ is
the critical density today, $c$ is the speed of light and $B$ is the size of the simulation box in units of $h^{-1}{\rm Mpc}$.

\tcr{For our $f(R)$ model and i}n code units,
Eq.~(\ref{poissonfr}) and Eq.~(\ref{frpoisson}) can be rewritten respectively as,
\begin{eqnarray}
\tilde{\nabla}^2\tilde{\phi}&=&2a\Omega_m^0(\tilde{\rho}-1)+\frac{a}{2}\Omega_m^0-\frac{a^4\Omega_m^0}{2}\left(\frac{Da^2}{\tilde{f}_R}\right)^{\frac{1}{p_+}}\nonumber\\
&+&2a^4(1-\alpha)\Omega_d^0,\label{codepoission}\\
\tilde{\nabla}^2\tilde{f}_R&=&-\frac{a\Omega_m^0}{\tilde{c}^2}(\tilde{\rho}-1)+\frac{a^4\Omega_m^0}{\tilde{c}^2}\left(\frac{Da^2}{\tilde{f}_R}\right)^{\frac{1}{p_+}}\nonumber\\
&-&\frac{4a^4(1-\alpha)\Omega_d^0}{\tilde{c}^2}-\frac{a\Omega_m^0}{\tilde{c}^2},\label{codefr}
\end{eqnarray}
where $\tilde{f}_{R}\equiv a^2 f_R$.

\begin{figure}
\includegraphics[width=3.5in,height=2.8in]{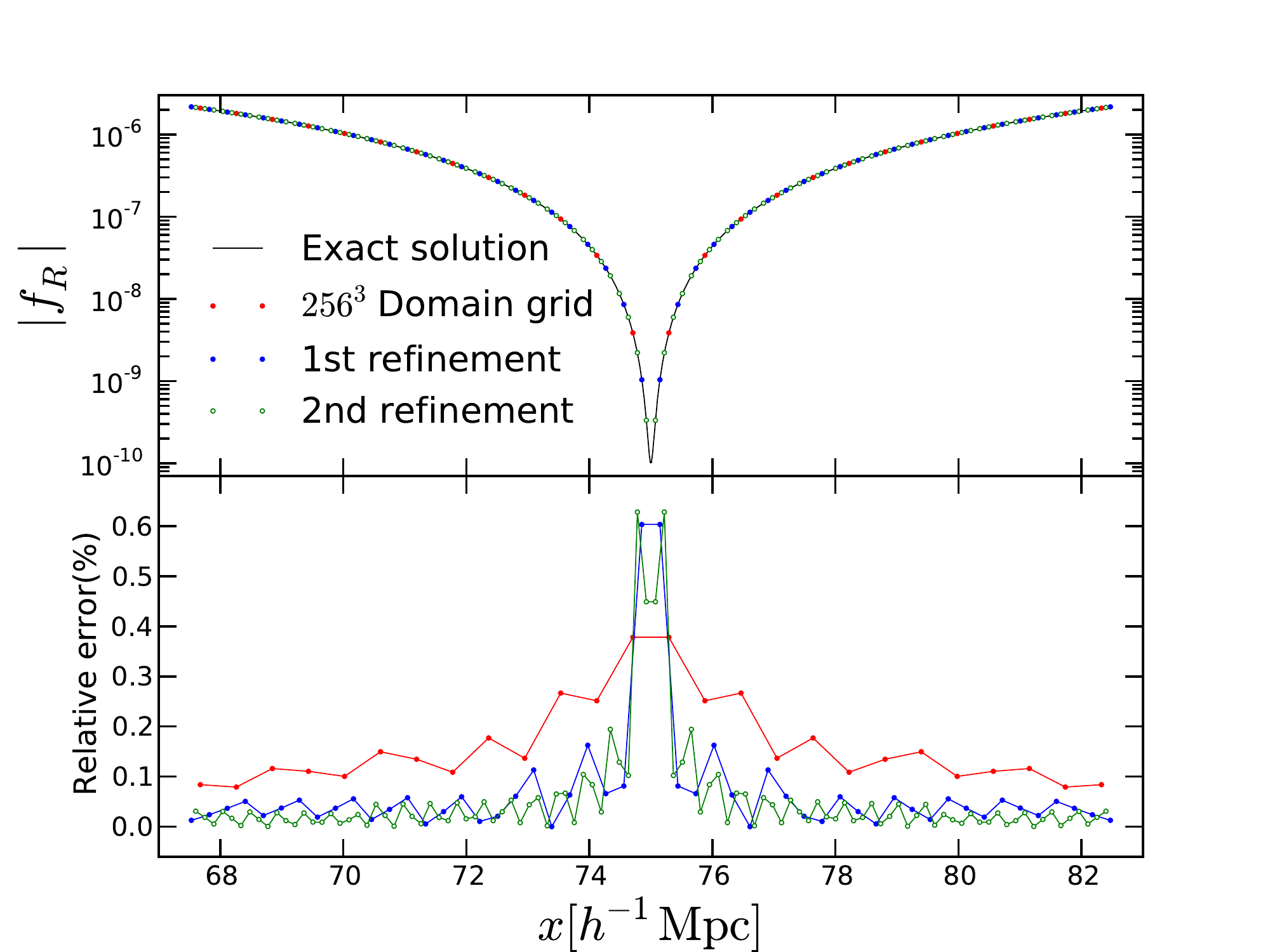}
\caption{Upper Panel:The numerical solution of the Gaussian field on the $256^3$ domain grids, as well as the first and second refinements. The solid line is the analytical solution. We take $|f_{R0}|=10^{-5}$ in the tests, and the size of the simulation box is $150h^{-1}{\rm Mpc}$. Lower Panel: The errors of the numerical results relative to the exact solution on the domain grid and each refinement.}\label{Gaussian_test}
\end{figure}

Since these equations are different from those in the default {\sc Ecosmog} code, we need to test the accuracy of our modified code.
Following~\cite{ECOSMOG}, we take the density $\delta$ as a one dimensional (in the $x$ direction without loss of generality) Gaussian field
\begin{equation}
\begin{split}
\delta(x)&=\left[\frac{(x-\frac{1}{2})^2}{W^2}-\frac{1}{2}\right]\frac{4\beta a \tilde{c}^2 \bar{f}_R(a)}{W^2\Omega_m^0}{\rm exp}\left(-\frac{\left(x-\frac{1}{2}\right )^2}{W^2}\right)\\
&+a^3\left(\frac{D}{\bar{f}_R(a)\left[1-\beta {\rm exp}\left(-\frac{\left(x-\frac{1}{2}\right )^2}{W^2}\right)\right]}\right)^{\frac{1}{p_+}}\\
&-4a^3(1-\alpha)\frac{\Omega_d^0}{\Omega_m^0}-1,
\end{split}
\end{equation}
which admits the following solution to the field $\tilde{f}_R$:
\begin{equation}
\tilde{f}_R(x)=a^2\bar{f}_R(a)\left[1-\beta {\rm exp}\left(-\frac{\left(x-\frac{1}{2}\right )^2}{W^2}\right)\right],
\end{equation}
where $W$ and $\beta$ are constants. We use $W=0.1$, $\beta=0.99999$ in the test. In Fig.~\ref{Gaussian_test}, we show the numerical results on domain grids, as well as the first and second refinements. The numerical results are in good agreement with the analytical solutions. In addition to the Gaussian field test, we have also tested the code with both sine and homogenous fields, and found the numerical results to be in excellent agreement with the analytical solutions. We will not present results of the latter tests here.

The perturbation equations in code units for the H-S model have been presented in Refs.~\cite{simulation,Zhao,ECOSMOG,Puchwein}. \tcr{Interested r}eaders are referred to these papers for further details\tcr{, and w}e will not repeat them here.

\subsection{Simulation details}

\begin{figure}
\includegraphics[width=3.5in,height=2.8in]{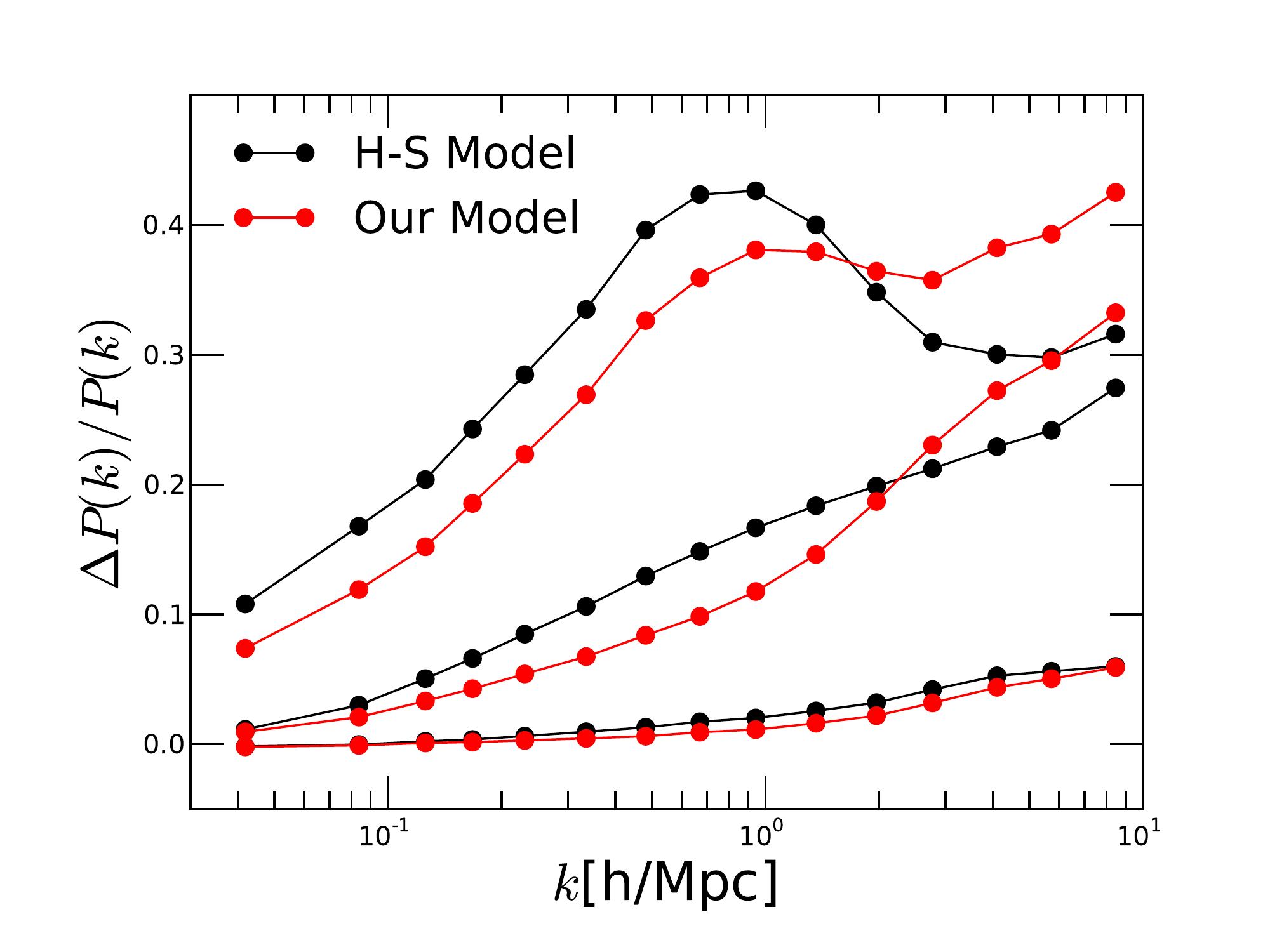}
\caption{The matter power spectra for our model (red) and the H-S (black) model measured from the simulations. From top to bottom, $f_{R0}$ takes the value $-10^{-4},-10^{-5},-10^{-6}$ respectively. }\label{deltPk}
\end{figure}

The cosmological parameters used in our simulations are $\Omega_b^0=0.049, \Omega_c^0=0.267, \Omega_d^0=0.684, h=0.671, n_s=0.962$, and $\sigma_8=0.834$, \tcr{which are} the Planck \cite{planck} best-fit values for \tcr{the standard} $\Lambda$CDM \tcr{model}. We use the {\sc Mpgrafic} package \cite{inicon} to generate initial conditions at $z_{\rm ini}=49$. The number of particles in our simulations is $N=256^3$ and the box size is $L_{\rm box}=150 h^{-1}{\rm Mpc}$. We run four realisations for each model. For each realisation, the different models share the same initial conditions. In Fig.~\ref{deltPk} we show the ratio of the power spectra $$\Delta P/P=P_{f(R)}(k)/P_{\Lambda \rm{CDM}}(k)-1$$ at $z=0$, measured using the {\sc powmes} \cite{POWMES} \tcr{code}. The power spectra are averaged over the four realisations.
The $f(R)$ parameter $f_{R0}$ is taken to be $f_{R0}=-10^{-6},-10^{-5},-10^{-4}$ for both our model and the H-S model. Compared with our previous work~\cite{Hesim}, we have significantly improved the accuracy of the background field $f_R$ in the regime $R\sim R_0$ by introducing the parameter $\alpha$ in the fitting formula Eq.~(\ref{eq:fitting}). When $\alpha=0$, the perturbation equations  Eq.~(\ref{codepoission}) and Eq.~(\ref{codefr}) reduce to the equations used in Ref.~\cite{Hesim}.

\section{Cosmological inequalities\label{special}}

In this section we will lay out the theoretical framework for the screening mechanism in $f(R)$ gravity. We will begin by discussing the importance of the homogenous field solution in $f(R)$ gravity and then introduce two inequalities. Using these inequalities, we will explain how the screening works. In the next section, we shall apply the theory presented here to dark matter haloes.

\subsection{Homogeneous density field}

We begin by discussing the solutions of Eq.~(\ref{poissonfr}) and Eq.~(\ref{frpoisson}) for a homogenous density field ($\delta \rho=0$). From Eq.~(\ref{frpoisson}), the vanishing \tcr{of} $\delta f_R$ gives
\begin{equation}
f_R=\bar{f}_{R}(\bar{R})=D\left(\frac{3\Omega_m^0H_0^2}{\bar{R}-12\alpha\Omega_d^0H_0^2}\right )^{p_+}, \label{app_background}
\end{equation}
where
\begin{equation}
\bar{R}(a)=[3\Omega_m^0a^{-3}+12\Omega_d^0]H_0^2.
\end{equation}
The error of the field $f_R$ obtained from Eq.~(\ref{app_background}) relative to the exact expression of the derivative of the background field Eq.~(\ref{fr}) is shown in Fig.~\ref{rel:lower}. As described above, the maximal deviation is \tcr{about} $5.5\%$ in the range $R_0<R<3.3R_0$ and, when $R>3.3R_0$, the error rapidly drops to below $1\%$. For the modified Poisson equation, Eq.~(\ref{poissonfr}), $\delta \rho=0$ gives the homogeneous solution of the field $\phi=0$, namely the zero point of the potential, which, as we shall show later, plays an important role in \tcr{understanding} the screening mechanism in $f(R)$ cosmology.

On the other hand, roughly speaking, when the local density in the simulations is above the background density  ($\rho>\bar{\rho}$), the potential $\phi$ is negative ($\phi<0$) and $\delta f_R$ is positive ($\delta f_R>0$). When the local density is below the background density  ($\rho<\bar{\rho}$), the potential $\phi$ is positive ($\phi>0$) and $\delta f_R$ is negative ($\delta f_R<0$). However, as we shall show later, the ratio
$-\frac{c^2\delta f_R}{\phi}$ is usually positive $-\frac{c^2\delta f_R}{\phi}>0$ because $\phi$ and $\delta f_R$ will change their signs simultaneously as $\phi$ crosses zero.

\subsection{Voids}

In this subsection, we will discuss solutions of the fields in void regions, \tcr{where} $\rho\sim0$.
In $f(R)$ gravity, voids are not really empty, but permeated with the scalar field $f_R$. The solutions of Eq.~(\ref{frpoisson}) in \tcr{these} regions are usually quite complicated \tcr{-- t}hey depend not only on the size of the void but also on the environment surrounding \tcr{it} \cite{ccl2013}. However, if we consider an extreme case where, for a large enough void, the distribution of the cosmic field $f_R$ near the void centre is nearly homogeneous ($\delta f_R\sim 0$), we have $\nabla^2\delta f_R\sim0$ and Eq.~(\ref{frpoisson}) yields
\begin{equation}
R\sim4\Lambda,\label{lowbound}
\end{equation}
where we have used the expression for the background Ricci curvature $\bar{R}$
\begin{equation}
\bar{R}=8\pi G \bar{\rho}+4\Lambda,
\end{equation}
and the assumption that at the void centre $\rho\sim0$ so that $\delta\rho\sim-\bar{\rho}$.

\begin{figure*}
\includegraphics[width=7in,height=4.8125in]{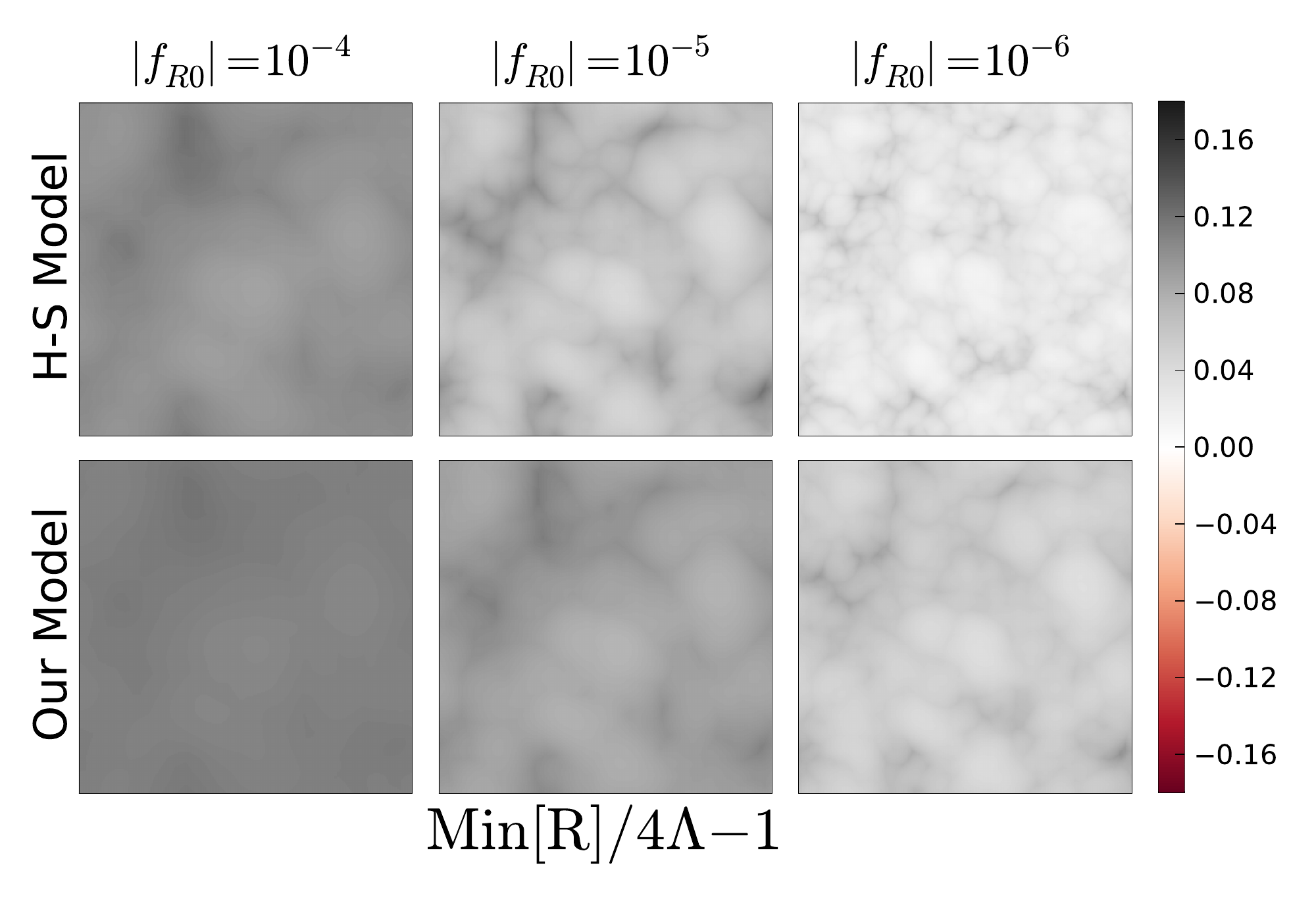}
\caption{The projected map of $\frac{{\rm Min}[R]}{4\Lambda}-1$ for the Hu-Sawicki model with $n=1$ \tcr{(upper row)} and our model \tcr{(lower row)}. We find the minimal value of $R$ along the $z$ direction in the simulation box \tcr{for each ($x$, $y$) point}.  In cases where $|f_{R0}|=10^{-4}$, the minimal values of $R$ are very close to $R_0$ \tcr{(the background curvature at present)} and the distribution of the projected value is close to homogenous. In cases where $|f_{R0}|=10^{-6}$, the minimal values of $R$ are very close to $4\Lambda$. We can see that $R$ is greater than $4\Lambda$ ($R>4\Lambda$) for both models. }\label{Rmap}
\end{figure*}

Eq.~(\ref{lowbound}) implies that in the perturbed Universe, even \tcr{at} the centres of voids, the local curvature $R$ in $f(R)$ gravity has a nonzero lower bound $4\Lambda$. This result does not assume any specific functional form of $f(R)$ and just requires that the background expansion is practically indistinguishable from that of the $\Lambda$CDM model. As a result, this conclusion is general. To \tcr{check} this explicitly, we generate a two dimensional map from our simulations by finding the minimal value of the curvature $R$ along the $z$ direction through the simulation box and project them onto the $x$-$y$ plane. As shown in Fig.~\ref{Rmap}, in the cases with $|f_{R0}|=10^{-6}$, the minimal values of $R$ are very close to $4\Lambda$, and we can see clearly that $R>4\Lambda$ for both $f(R)$ models. In the cases with $|f_{R0}|=10^{-4}$, the minimal values of $R$ are very close to $R_0$ and the distribution of $\rm{Min}[R]$ is nearly homogeneous. These numerical \tcr{checks} thus confirm that
\begin{equation}
R>4\Lambda.
\end{equation}

From this inequality, we know that the approximate formulae for the background fields $f_R$ (e.g.,~Eq.~(\ref{fR:approx}) and Eq.~(\ref{appHI})) only need to be accurate in the range $R>4\Lambda$.  Furthermore, $f(R=0)=0$ is not a necessary condition for $f(R)$ models, given the fact that the point $R=0$ will never be arrived at in the Universe since $R>4\Lambda$ if the background expansion of the $f(R)$ model is practically indistinguishable from the $\Lambda$CDM model. Nevertheless, our model explicitly predicts $R>4\Lambda$ and is therefore naturally consistent with this inequality.

\subsection{High density regions}

In this subsection, we will discuss the solutions of Eqs.~(\ref{poissonfr}, \ref{frpoisson}) in regions of high density. There are two types of solutions. If $\delta R \approx 8\pi G\delta \rho$, the solution is called the high-curvature solution. Correspondingly, the solution with $\delta R \ll 8\pi G\delta\rho$ is called the low-curvature solution. Note that high density does not necessarily imply high curvature in $f(R)$ gravity.

The low-curvature solution is usually arrived at when the amplitude of the background field, $|\bar{f}_R|$, is large compared to the local potential: $c^2|\bar{f}_R|\tcg{>}|\phi|$~\cite{simulation}. The terms which are associated with the perturbation of the curvature, $\delta R(f_R)=\frac{\partial R}{\partial f_R}\delta f_{R} \ll 8\pi G\delta \rho$, in Eqs.~(\ref{poissonfr}) and (\ref{frpoisson}) have a minor effect and can be neglected. These equations can therefore be linearised and reduced to
\begin{eqnarray}
\nabla^2\phi&\approx&\frac{16\pi G}{3}\delta \rho\label{linpossion}\\
\nabla^2\delta f_R&\approx&-\frac{8\pi G}{3c^2}\delta \rho\label{linfr}.
\end{eqnarray}
Eqs.~(\ref{linpossion}, \ref{linfr}) indicate that, given the density field $\delta \rho$ and under the same (e.g., periodic) boundary conditions, their solutions 
satisfy the relation $c^2\delta f_R\sim -{\phi}/{2}$.
In this extreme case, the scalar field $|\delta f_R|$ and the local potential $|\phi|$  attain their maximum values as $|-\frac{2\phi_N}{3}|$ and $|-\frac{4\phi_N}{3}|$ respectively, where $\phi_N$ is the standard Newtonian potential for the given density field $\delta \rho$. Combining Eq.~(\ref{poissonfr}) and Eq.~(\ref{frpoisson}), we obtain
\begin{equation}
\nabla^2\left(\phi+\frac{c^2\delta f_R}{2}\right)=4\pi G\delta \rho \label{poissonN}.
\end{equation}
The standard Newtonian potential, $\phi_N$, \tcr{is related} to the total potential $\phi$ and the scalar field $c^2\delta f_R$ as
\begin{equation}
\phi_N=\phi+\frac{c^2\delta f_R}{2}.\label{Nphi}
\end{equation}

In general, if the background field $|\bar{f}_R|$ is not large enough, we have
\begin{equation}
 c^2\left|\delta f_R\right|\leq \left|-\frac{2\phi_N}{3}\right|,\label{thin_shell}
\end{equation}
which is \tcr{a known result} in the literature \cite{HuS,Khoury}. Furthermore, in high-density regions, we usually have $\phi_N<0$, $\phi<0$ and $\delta f_R>0$. Inserting Eq.~(\ref{Nphi}) into Eq.~(\ref{thin_shell}), we have
\begin{equation}
c^2\left|\delta f_R\right| \leq \left|-\frac{\phi}{2}\right|,\label{strongthinshell}
\end{equation}
which only involves the quantities $\delta f_R$ and $\phi$ (remember that $\phi_N$ is not a physical quantity in $f(R)$ gravity).
In high-density regions, 
applying Eq.~(\ref{Nphi}) and Eq.~(\ref{strongthinshell}), and using $\phi_N<0$, $\phi<0$, $\delta f_R>0$, we obtain
\begin{equation}
\left|-\phi_N\right| \leq \left|-\phi\right| \leq \left|-\frac{4}{3}\phi_N\right|\label{phi_phi},
\end{equation}
where the left and right limits correspond to \tcr{the extreme cases of} high-curvature and low-curvature solutions, respectively.
It is \tcr{evident} that Eq.~(\ref{phi_phi}) is equivalent to the well-known result that $G\leq G_{\rm eff}\leq\frac{4}{3}G$ in $f(R)$ gravity, where $G_{\rm eff}$ is the effective Newtonian constant which is defined by
\begin{equation}
\frac{G_{\rm eff}}{G}\equiv\frac{4}{3}-\frac{\delta R}{3\kappa^2\delta \rho}.\label{geff}
\end{equation}
$G_{\rm eff}$ determines the strength of the gravitational interactions between massive particles in $f(R)$ gravity and $G$, on the other hand, is what is felt by photons and other massless particles.

From Eq~(\ref{phi_phi}),
we notice that Eq.~(\ref{strongthinshell}) imposes a tighter constraint on the scalar field perturbation $c^2|\delta f_R|$ than
Eq.~(\ref{thin_shell}). We therefore will focus on Eq.~(\ref{strongthinshell}) throughout this work\tcr{, and take it as} the starting point of our analyses for the the rest of this paper. 
We will first check its validity against our numerical simulations, before trying to quantitatively understand the screening mechanism in $f(R)$ gravity \tcr{based} on it.

\begin{figure*}
\includegraphics[width=3.5in,height=2.8in]{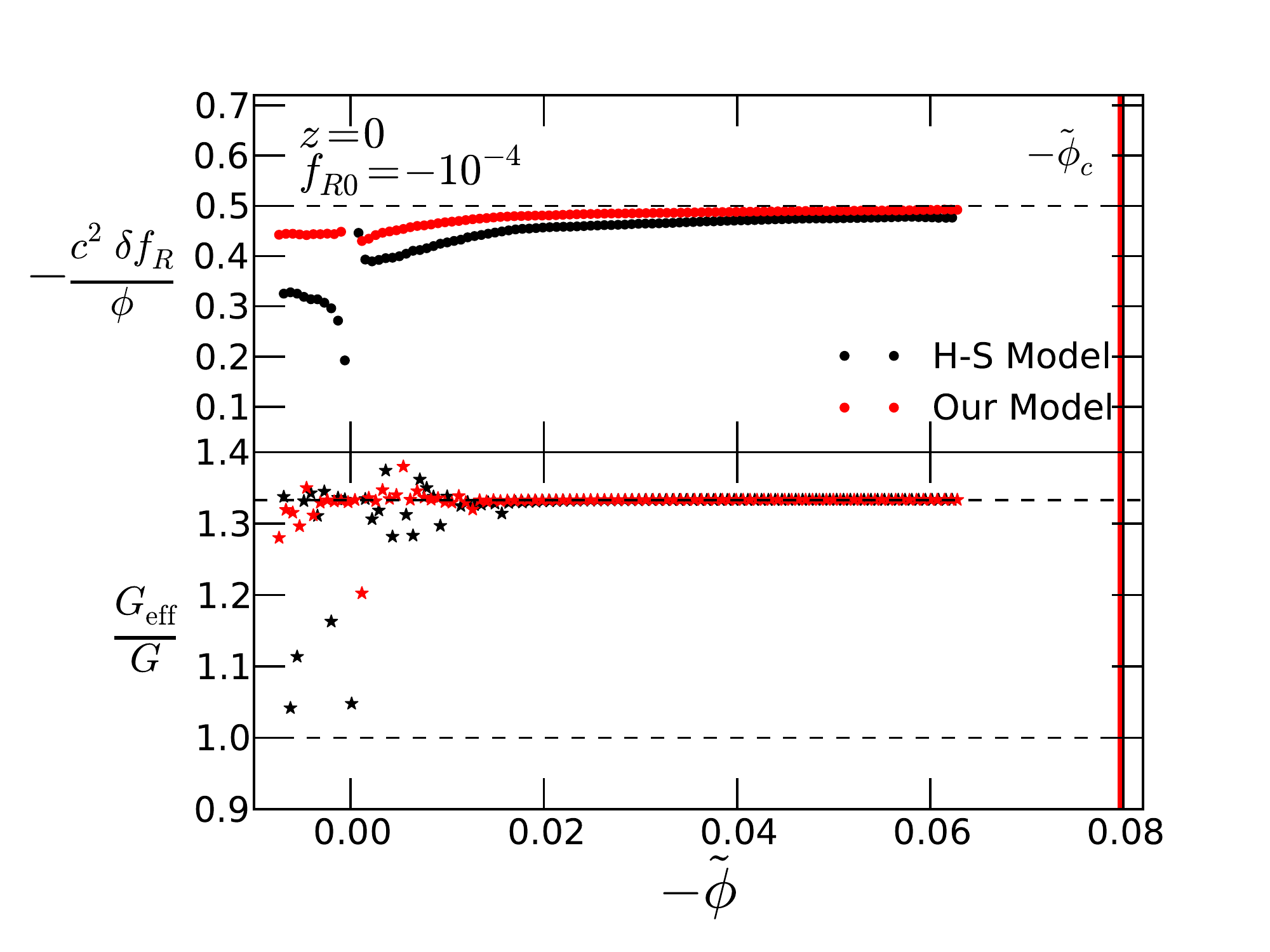}
\includegraphics[width=3.5in,height=2.8in]{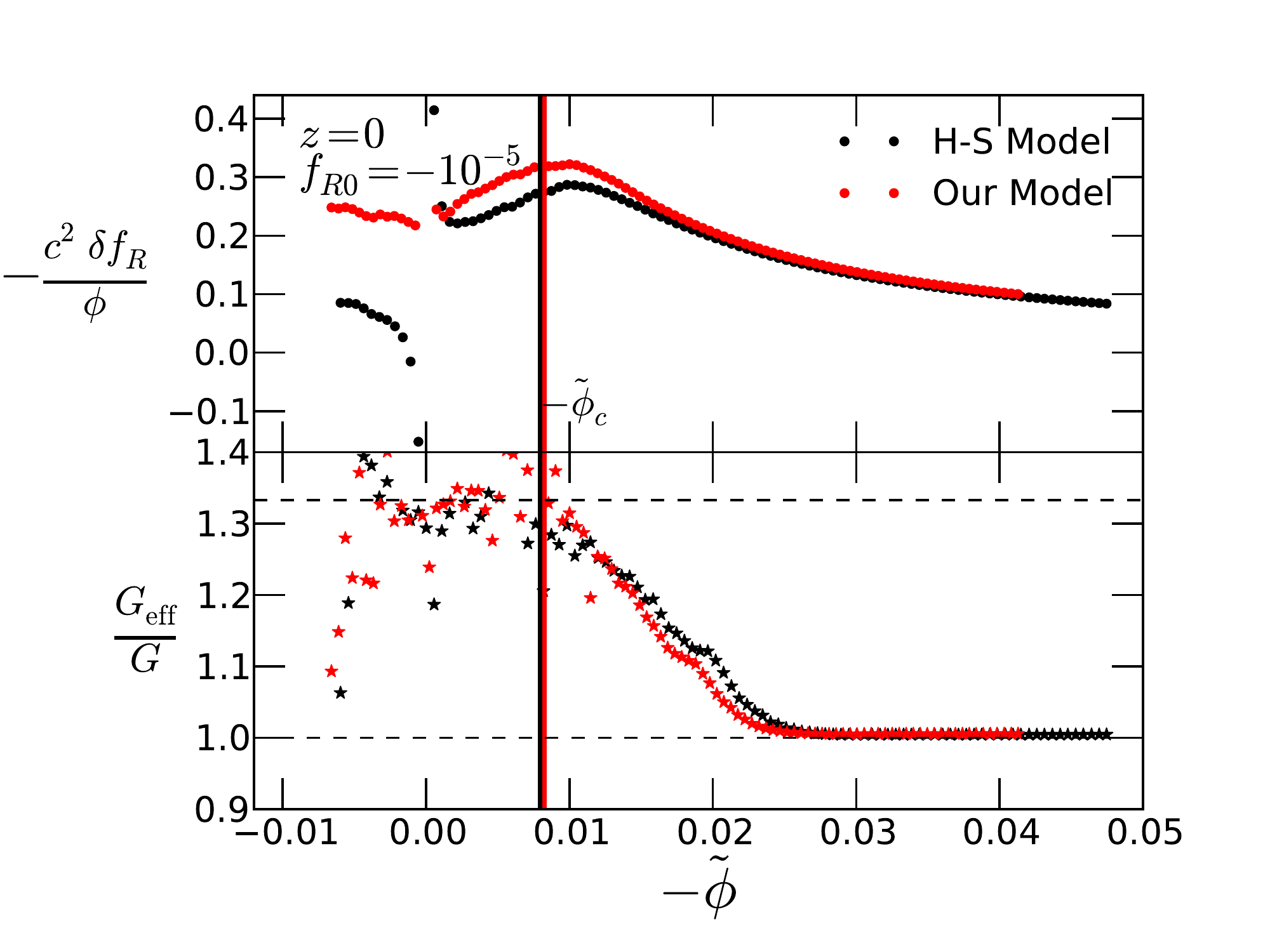}
\includegraphics[width=3.5in,height=2.8in]{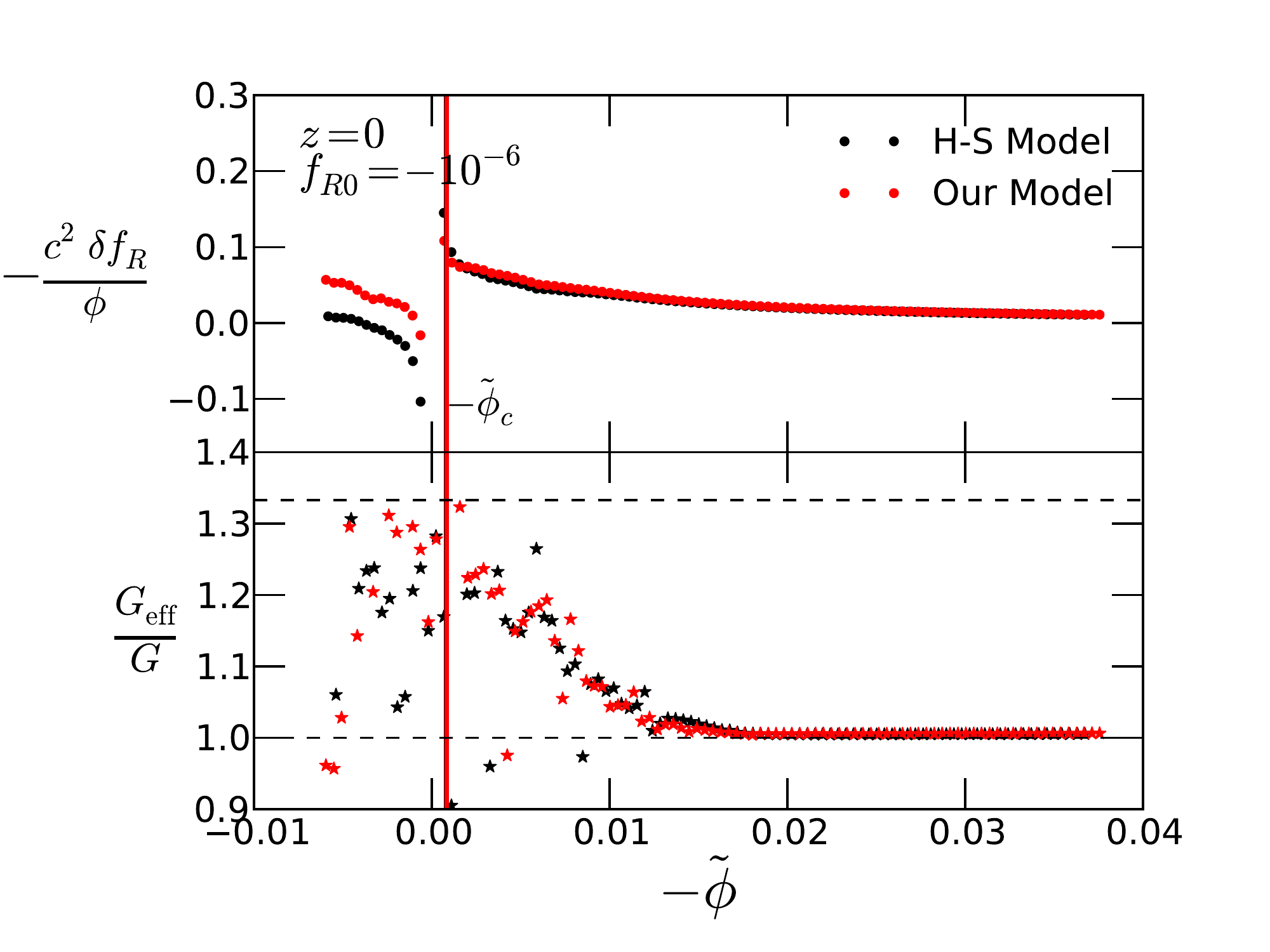}
\caption{ The statistics of  $-\frac{c^2\delta f_R}{\phi}$ and $\frac{G_{\rm eff}}{G}$ for our model and the H-S model at $z=0$. The \tcr{horizontal axis is the} potential $\phi$ in code units. The condition $|\phi|>2c^2|\bar{f}_{R}|$ is equivalent to $|\tilde{\phi}|>|\tilde{\phi}_c|$ where  $\tilde{\phi}_c=2\tilde{c}^2\bar{f}_{R}$ \tcr{is the critical potential and is indicated by red (black) solid vertical lines for our (the H-S) model}. 
We can see clearly that when $|\tilde{\phi}|>|\tilde{\phi}_c|$, the screening mechanism starts to work. }\label{Fscreening}
\end{figure*}

\begin{figure*}
\includegraphics[width=3.5in,height=2.8in]{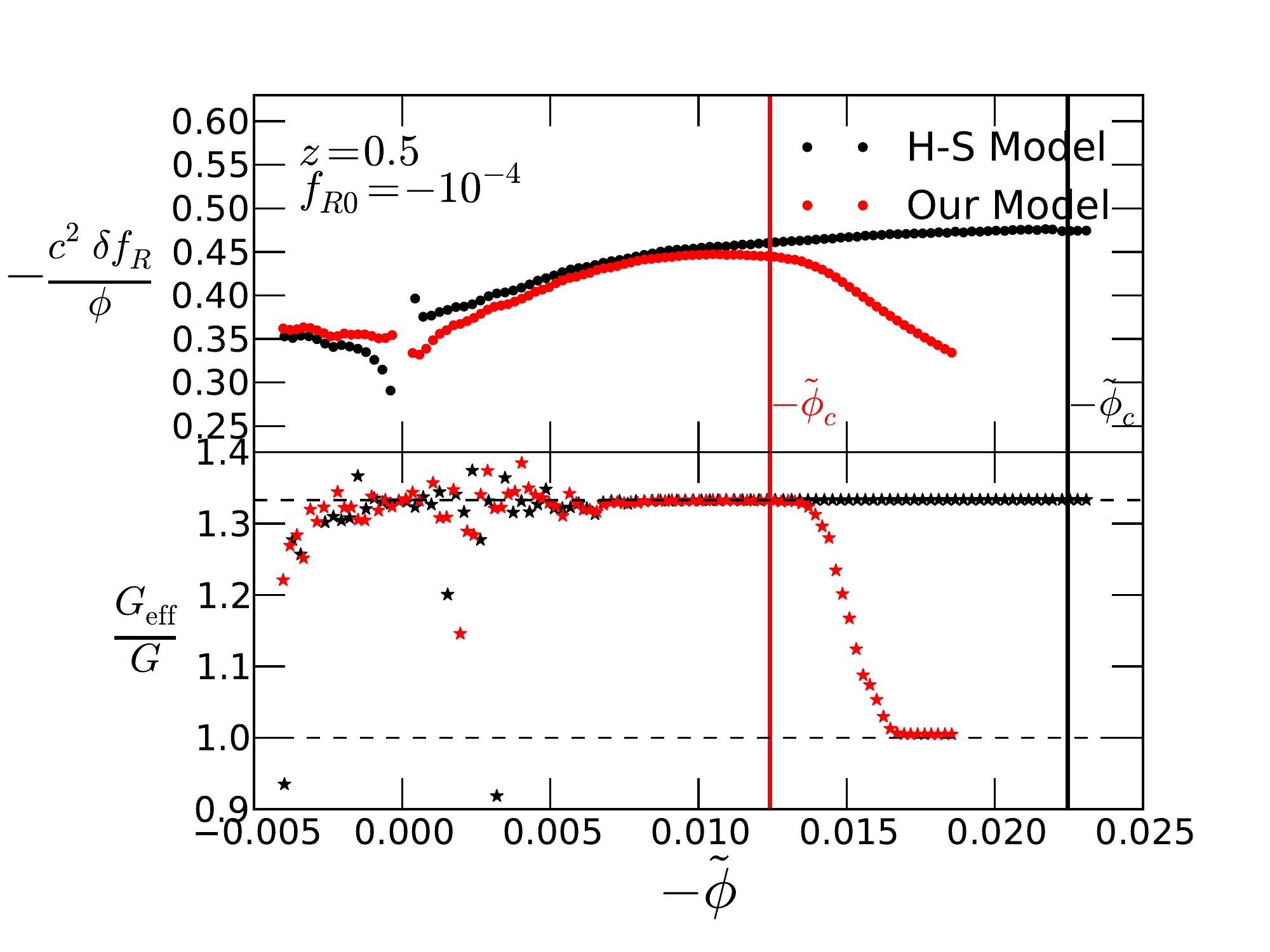}
\includegraphics[width=3.5in,height=2.8in]{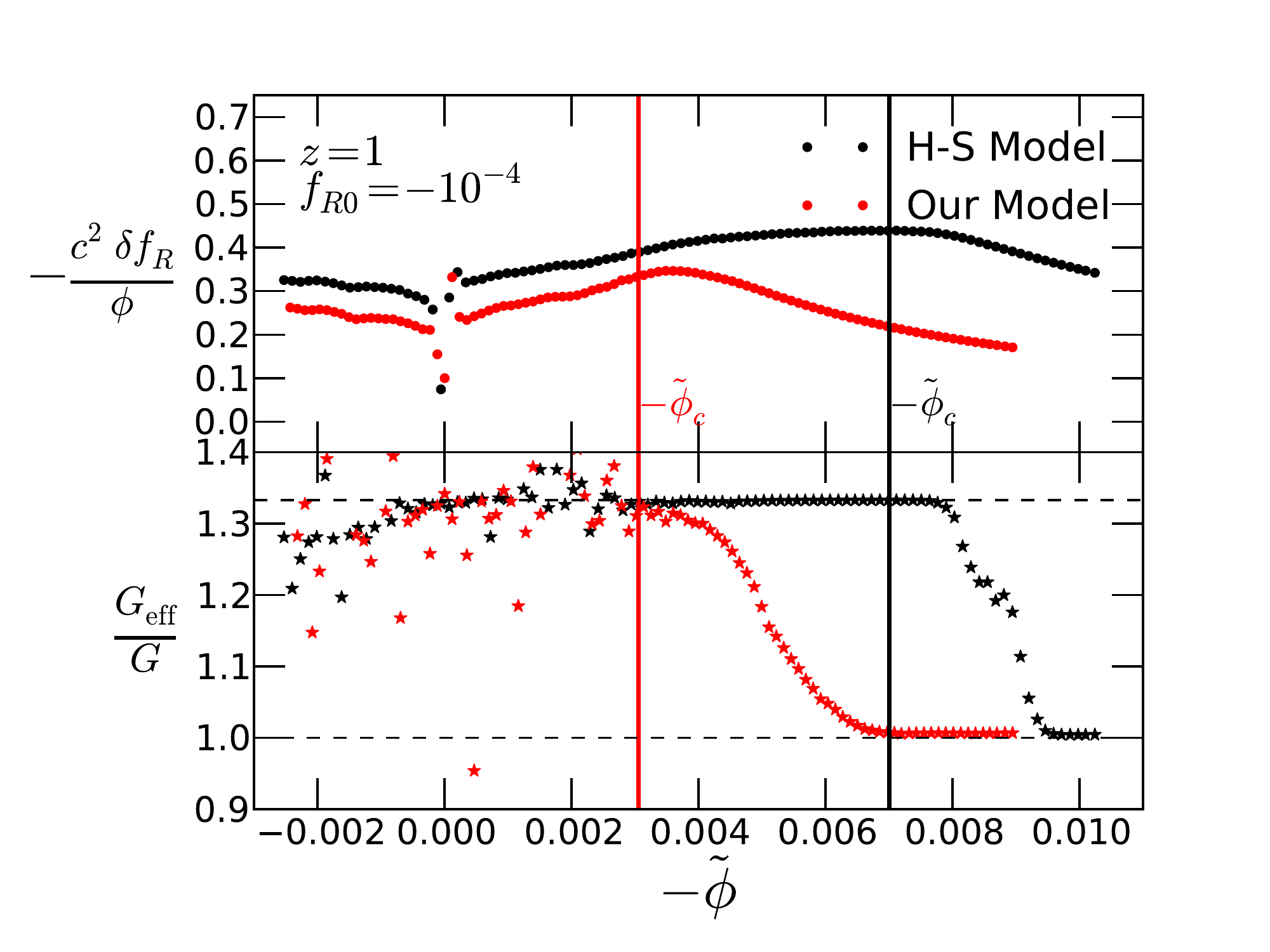}
\includegraphics[width=3.5in,height=2.8in]{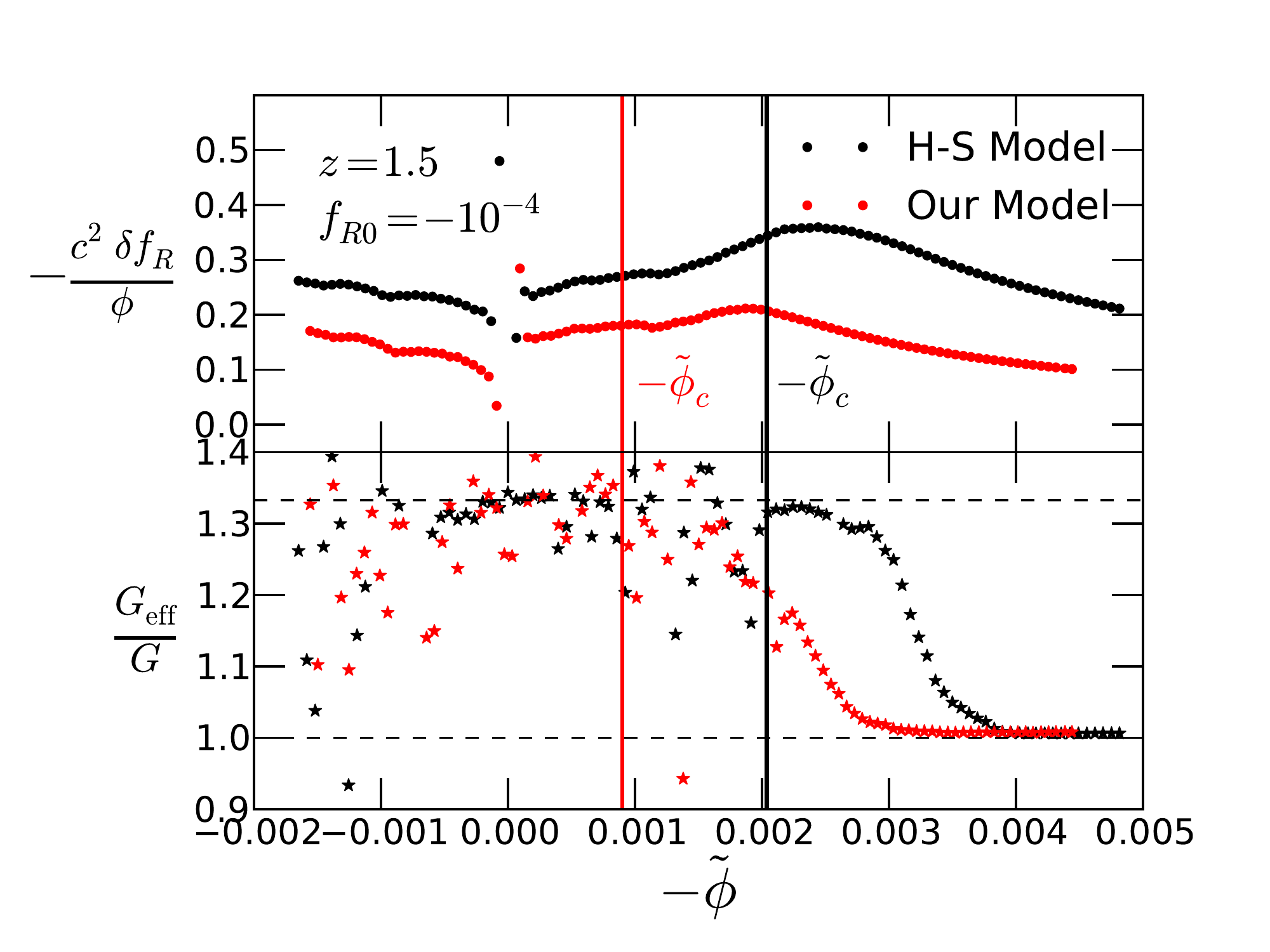}
\includegraphics[width=3.5in,height=2.8in]{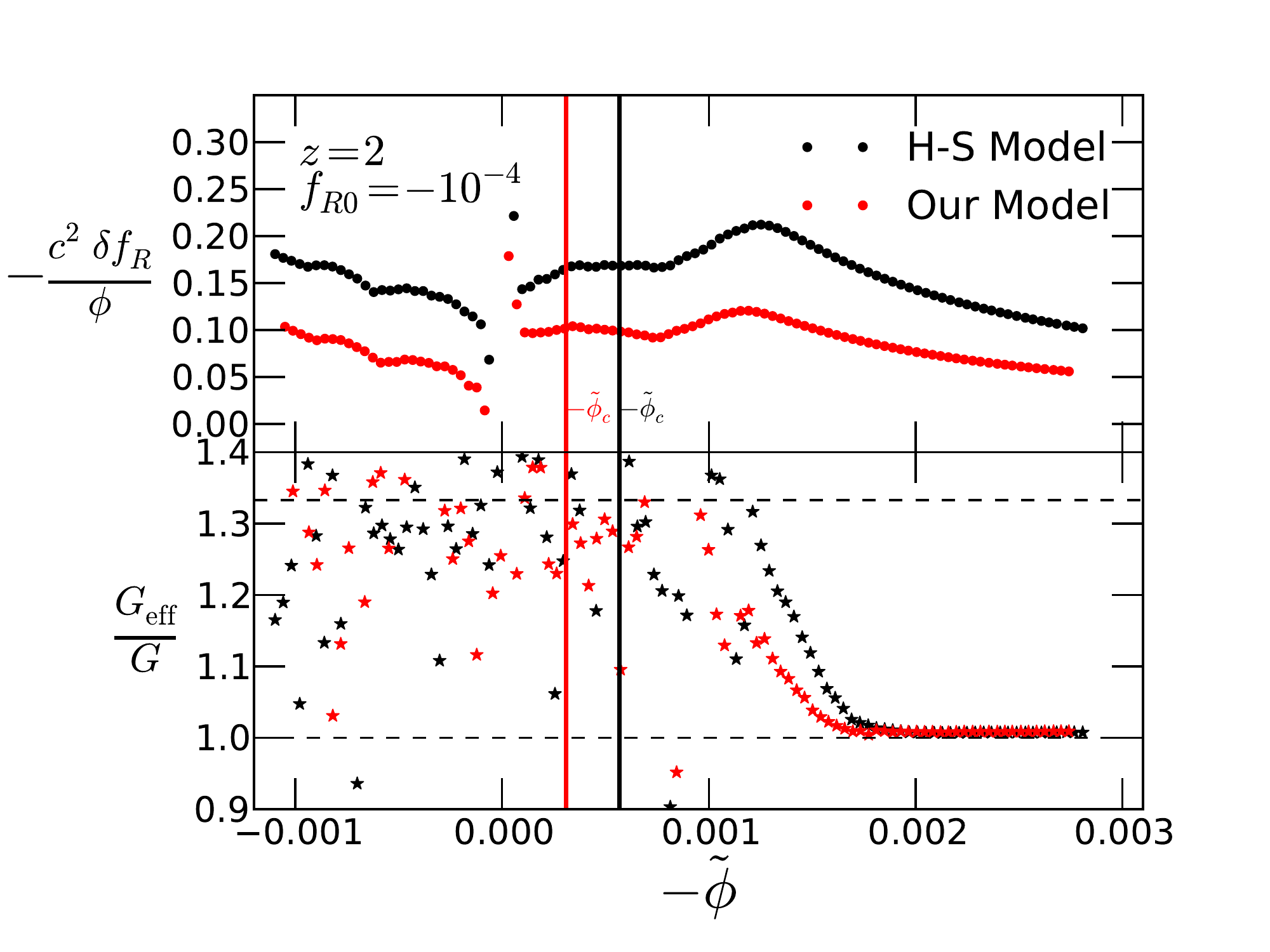}
\caption{The statistics of  $-\frac{c^2\delta f_R}{\phi}$ and $\frac{G_{\rm eff}}{G}$ for our model and the H-S model at high\tcr{er} redshifts. We take $f_{R0}=-10^{-4}$ for illustrative purpose. The potential $\phi(z)$ is in code units. The condition $|\phi(z)|>2c^2|\bar{f}_{R}(z)|$ is equivalent to $|\tilde{\phi}(z)|>|\tilde{\phi}_c(z)|$ where  $\tilde{\phi}_c(z)=2\tilde{c}^2\bar{f}_{R}(z)/(1+z)^2$. The red and black solid vertical lines indicate the critical values $\tilde{\phi}_c(z)$ for our model and the H-S model respectively. When $|\tilde{\phi}(z)|>|\tilde{\phi}_c(z)|$, the screen mechanism starts to work.}\label{screening}
\end{figure*}

To check Eq.~(\ref{strongthinshell})
in our simulations, we statistically compare the values of $-\frac{c^2\delta f_R}{\phi}$ and $-\phi$. We divide the potential $\phi$ into $100$ equal bins from the minimal value to the maximal value. For convenience, $\phi$ is in code units. We then count the number of occurrences of $-\frac{c^2\delta f_R}{\phi}$
and calculate its arithmetic average in each bin.
The results are shown in the upper panels in each plot of Fig.~\ref{Fscreening}. Included in Fig.~\ref{Fscreening} are the results at $z=0$ for our $f(R)$ model (red) and for the H-S model (black), each with different parameters $f_{R0}=-10^{-4},-10^{-5},-10^{-6}$. We clearly find there that
$-\frac{c^2\delta f_R}{\phi}$ is a positive and rather smooth function with respect to the potential $\phi$,
except in the vicinity of $\phi=0$, where the \tcr{discontinuities} are due to numerical errors. We find that the maximal value of $-\frac{c^2\delta f_R}{\phi}$ is $1/2$, which only happens in the $f_{R0}=-10^{-4}$ case. In the other two cases  ($f_{R0}=-10^{-5},-10^{-6}$), the value of $-\frac{c^2\delta f_R}{\phi}$ is much smaller than $0.5$. Our numeric simulations therefore confirm Eq.~(\ref{strongthinshell}). For completeness, we also check this issue at higher redshifts ($z=0.5,1,1.5,2$). \tcr{Taking} $f_{R0}=-10^{-4}$ as an example, as shown in the upper panels in each plot of Fig.~\ref{screening}, Eq.~(\ref{strongthinshell}) also holds at higher redshifts.

We are now in a position to understand the screening mechanism using Eq.~(\ref{strongthinshell}).
We shall focus on high-density regions ($\delta \gg 1$) in this work. 
As mentioned above, in these regions, the potential is usually negative ($\phi<0$) and the \tcr{magnitude} of the scalar field $f_R$ \tcr{smaller} than the value of the background field, $|f_R|<|\bar{f}_R|$  (see Fig.~\ref{screening}), implying that $\delta f_R>0$. Equation~(\ref{strongthinshell}) in this case can be rewritten as
\begin{equation}
-\frac{\phi}{2}\geq c^2(f_R-\bar{f}_R)=c^2\delta f_R>0\label{ineq},
\end{equation}
from which we have
\begin{equation}
c^2f_R\leq-\frac{\phi}{2}+c^2\bar{f}_R.\label{ineq2}
\end{equation}
Recall that $f_R$ must satisfy the physical constraint $f_R<0$ due to the stability considerations of the \tcr{perturbation evolution} 
in the high curvature regime~\cite{Ignacy}, it can be shown that
{if the right hand side of Eq.~(\ref{ineq2}) is less than zero or,
equally, $c^2\bar{f}_R<\frac{\phi}{2}$}, the absolute value of $c^2f_R$ will have a nonzero lower bound: $c^2|f_R|\geq|-\frac{\phi}{2}+c^2\bar{f}_R|>0$. If the background field $|\bar{f}_R|$ is large ($c^2|\bar{f}_R|\gg |\frac{\phi}{2}|$), this lower bound will be rather high as well ($|-\frac{\phi}{2}+c^2\bar{f}_R|\gg 0$), which means that $|f_R|$ can not be adequately suppressed in high-density regions, leading to a strong fifth force. This physical picture can also be viewed in a different way: the existence of the lower bound for $|f_R|$, for both $f(R)$ models studied in this work, conversely,  means that there is
an upper bound on the curvature: $R_{\rm max}=R(f_R=-|-\frac{\phi}{2c^2}+\bar{f}_R|)$ in high-density regions. If $R_{\rm max}\ll 8\pi G \rho$, the solution to the curvature is far below the GR prediction, \tcr{so that} the model does not have a high-curvature solution in high-density regions and would be ruled out. Therefore, $c^2|\bar{f}_R|\gg |-\frac{\phi}{2}|$ is a sufficient condition for the model to admit the low-curvature solution.

On the other hand, if $-\frac{\phi}{2}+c^2\bar{f}_R\sim0$, the magnitude of the scalar field $f_R$ can be sufficiently suppressed:
$|f_R|\rightarrow0$ and $R_{\rm max}$ can be close enough to its GR solution, $R_{\rm max}\sim 8\pi G\rho$, so that a $f(R)$ model could admit the high-curvature solution. Moreover, if the local scalar field $\phi$ satisfies $|\phi|>2c^2|\bar{f}_R|$, there will be no constraint on the maximal value of the local scalar curvature ($R_{\rm max}=+\infty$), and the high-curvature solution can possibly be arrived at too. $|\phi|\gtrsim2c^2|\bar{f}_R|$ is therefore the necessary condition for the high-curvature solution. However, 
\tcr{this is not a sufficient condition:} as we shall show later, to guarantee a high-curvature solution ($G_{\rm eff}\sim G$), the potential well $\phi$ need to be deep enough relative to the background field $2c^2|\bar{f}_R|$.

In order to test the above conclusions, we perform a similar statistical analysis, to that of $-\frac{c^2\delta f_R}{\phi}$, for the effective Newtonian constant $G_{\rm eff}$, which is defined by Eq.~(\ref{geff}). Recall that $G_{\rm eff}\sim G$
indicates the high-curvature solution ($\delta R\sim \kappa^2\delta \rho$) and
$G_{\rm eff}\sim \frac{4}{3}G$
implies the low-curvature solution ($\delta R\ll\kappa^2\delta \rho$). The numerical results for the statistics of $G_{\rm eff}/G$ are shown in the lower panels in each plot of Fig.~\ref{Fscreening} and Fig.~\ref{screening}. We define a critical potential as $\phi_c=2c^2\bar{f}_R$, and in Fig.~\ref{Fscreening} and Fig.~\ref{screening} $\phi_c$ (in code units) is indicated by vertical lines. As we have expected, when the magnitude of the local potential $|\phi|$ is higher than the critical potential $|\phi_c|$, the screening mechanism
starts to work, as can be seen clearly in Fig.~\ref{Fscreening} for both
$f(R)$ models studied, and for different values of the parameter $f_{R0}$.
For completeness, we also check this conclusion at higher redshifts ($z=0.5,1,1.5,2$). We take $f_{R0}=-10^{-4}$ for illustration purposes. Fig.~\ref{screening} shows that $|\phi_c|$ lies accurately at the point above which the screening mechanism starts to work. These numerical results are in good agreement with our above analysis. From Fig.~\ref{Fscreening} and Fig.~\ref{screening}, we can also see that high-curvature solutions with an effective Newtonian constant close to that in standard gravity, $G_{\rm eff}\approx G$, usually happen in regimes where the potential $\phi$ is substantially deeper than $\phi_c$.

Before leaving this section, we briefly summarise the main results obtained from the above analyses:
\begin{itemize}
\item $2c^2|\bar{f}_R|\gg |-\phi|$ is a sufficient condition for the low-curvature solution. Combining the constraint $R>4\Lambda$ obtained above, we find that the curvature scalar $R$ is bounded locally as
\begin{equation}
4\Lambda<R<R\left(f_R=-|-\frac{\phi}{2c^2}+\bar{f}_R|\right),\nonumber
\end{equation}
for the low-curvature solution. 
\tcr{If this occurs in the Solar system,} the model is ruled out.
Using $|-\frac{4}{3}\phi_N |\geq|-\phi|$, it
can be shown that $2c^2|\bar{f}_R|\gg |-\frac{4}{3}\phi_N|$ is also a sufficient condition for the low-curvature solution and is indeed stronger than the condition $2c^2|\bar{f}_R|\gg |-\phi|$ because, logically, we have
\begin{equation}
2c^2|\bar{f}_R|\gg |-\frac{4}{3}\phi_N|\ \Rightarrow\ 2c^2|\bar{f}_R|\gg |-\phi|.\nonumber
\end{equation}
\item $|-\phi|\gtrsim2c^2|\bar{f}_R|$ is a necessary but not sufficient condition for the high-curvature solution. From $|-\frac{4}{3}\phi_N |\geq|-\phi|$, we can show that $|-\frac{4}{3}\phi_N|\gtrsim2c^2|\bar{f}_R|$  is also a necessary condition for the high-curvature solutions. However, it is much weaker than that of $|-\phi|\gtrsim2c^2|\bar{f}_R|$ because, logically, we have
    \begin{equation}
    |-\phi|\gtrsim2c^2|\bar{f}_R|\ \Rightarrow\ |-\frac{4}{3}\phi_N|\gtrsim2c^2|\bar{f}_R|.
    \end{equation}
\end{itemize}

In addition to the above results, we also find that the critical potential $\phi_c=2c^2\bar{f}_R$ is a good indicator which tells us when the screening mechanism starts to work.
Such a universal criterion applies excellently to both $f(R)$ models studied here,
with different parameters ($f_{R0}=-10^{-6}, -10^{-5}, -10^{-4}$) at different redshifts (see Fig.~\ref{screening}). Regions where the local potential $\phi$ is below
$|\phi_c|$ are usually completely unscreened.

A {\it potential} application of the result obtained above is that the condition $|\frac{4}{3}\phi_N|<|\phi_c|$ can be used to identify unscreened galaxies and to make 
screening maps for galaxy surveys~\cite{cabre}. Such maps play an important role in astrophysical constraints on $f(R)$ gravity~\cite{Jain}, which can place much tighter constraint than what can be obtained from cosmological observations.

Nevertheless, there are some caveats before applying the conclusions made in this section to real galaxies. In the widely accepted picture, galaxies often form inside dark matter haloes, which are highly biased tracers of the underlying dark matter field. To make the necessary connections, we will extend our analysis to dark matter haloes in the next section.

\section{Dark matter haloes\label{haloes}}

From the previous analysis, we know that the screening in $f(R)$ gravity depends mainly on the depth of the gravitational potential. From the condition $|-\phi|\gtrsim2c^2|\bar{f}_R|$, we can infer that there are two possible ways for a dark matter halo to be screened. Firstly, the halo itself is so massive that it can generate a deep enough potential well that satisfies $|-\phi|\gg|-\phi_c|$:
this case is dubbed \emph{self-screening} ~\cite{cabre,hui,voids,Zhao_envir}. Secondly, for a halo too small to be self-screened but lying in a very deep potential well, if the magnitude of the total local potential satisfies $|-\phi|\gg|-\phi_c|$, then the halo can still become screened: this case is called \emph{environmental-screening}~\cite{cabre,hui,voids,Zhao_envir}. In the following, we will discuss these two different screening scenarios 
in detail.

We identify haloes in our simulations using a modified version of the {\sc AHF} code~\cite{AHF}. We follow the standard procedure in the {\sc AHF} code to locate density peaks as the positions of the dark matter haloes, but remove the unbound particles in haloes by taking into account the modification to gravity. We use the effective density $\delta \rho_{\rm eff}\equiv\frac{G_{\rm eff}}{G}\delta \rho$ instead of $\delta \rho$ to calculate the gravitational potential. In order to characterise screened and unscreened dark matter haloes, we follow~\cite{Zhao_envir} by defining the lensing mass $M_{L}$ and dynamical mass $M_{D}$ for a dark matter halo.

The lensing mass is the \tcr{bare} mass of the dark matter haloes, which is defined by
\begin{equation}
M_{\rm L}=\int \delta \rho({\bf x}) \,dV.
\end{equation}
The dynamic mass, on the other hand, is defined by
\begin{equation}
M_{\rm D}=\int \delta \rho_{\rm eff}({\bf x}) \, dV,
\end{equation}
which includes the effect of the scalar field. For a totally unscreened halo, the ratio between the two masses is $\frac{M_{D}}{M_{L}}\approx \frac{4}{3}$, while for a well screened 
halo we have $\frac{M_{D}}{M_{L}}\approx 1$. \tcr{In general, however, the value of $\frac{M_{D}}{M_{L}}$ is somewhere in between.}

\begin{figure*}
\includegraphics[width=6.5in,height=2.7in]{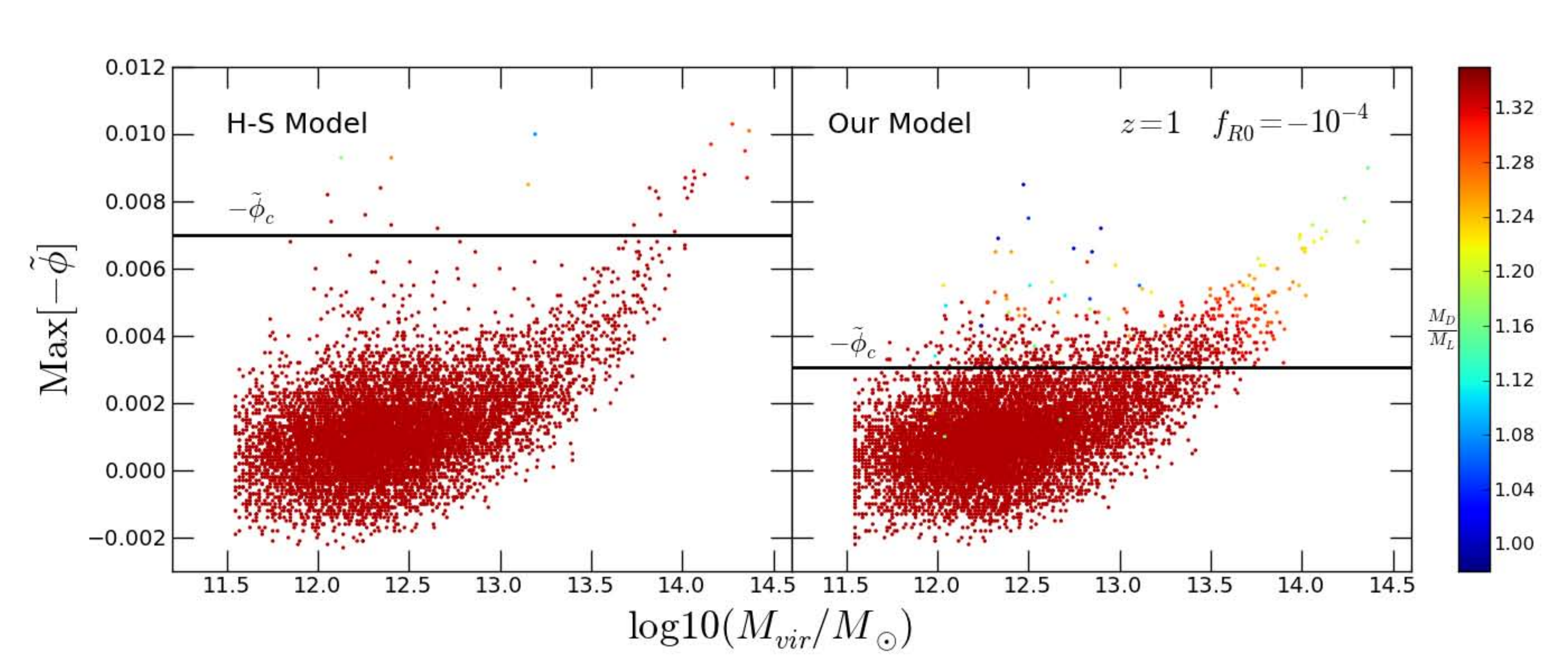}
\caption{Scatter plot for the maximal value of the gravitational potential $\rm Max[-\phi]$ inside a dark matter halo with respect to the lensing mass of the halo for $f(R)$ models with $f_{R0}=-10^{-4}$ at $z=1$. Each point represents a dark matter halo and its color \tcr{encodes} the ratio between the dynamical mass and the lensing mass 
\tcr{(see the colorbar on the right hand side)}. $|\tilde{\phi}_c|=2\tilde{c}^2|\bar{f}_R|$ is the critical value in code units, above which the screening mechanism starts to work. Halos with the maximal depth of the potential well $|-\phi|$ below the threshold $|\tilde{\phi}_c|$ are completely unscreened in this case. On the right panel, some small haloes are well screened due to environmental screening. However, in this case large haloes cannot generate deep enough potential wells \tcr{for self-screening} and therefore are only partially screened. }\label{F4halo}
\end{figure*}
\begin{figure*}
\includegraphics[width=6.5in,height=2.7in]{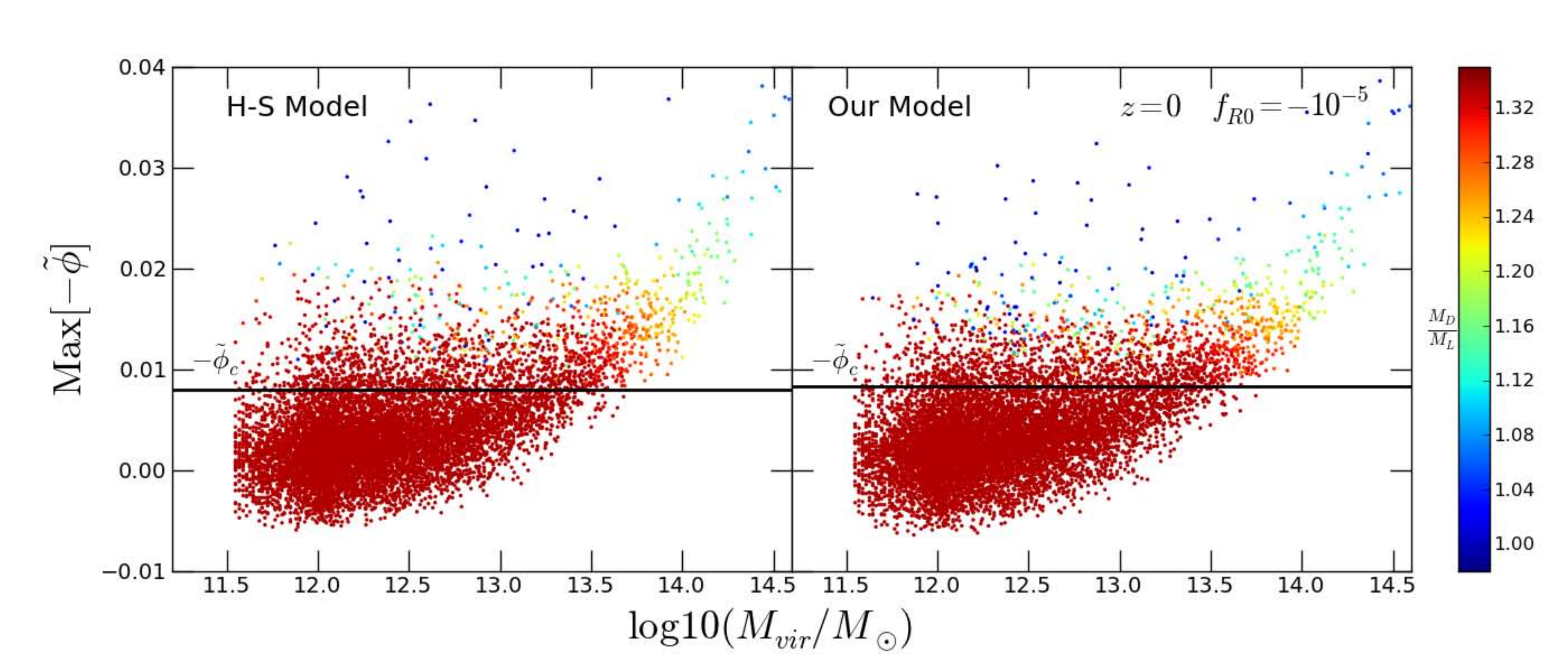}
\caption{Scatter plot for the maximal value of the gravitational potential $\rm Max[-\phi]$ inside a dark matter halo with respect to the lensing mass of the halo for $f(R)$ models with $f_{R0}=-10^{-5}$ at $z=0$. It is clear that below the horizontal line\tcr{, which represents} the critical potential $|\tilde{\phi}_c|=2\tilde{c}^2|\bar{f}_R|$, the haloes are completely unscreened. It is also clear that most of the well-screened haloes lie in very deep potential wells.
 }\label{F5halo}
\end{figure*}
\begin{figure*}
\includegraphics[width=6.5in,height=2.7in]{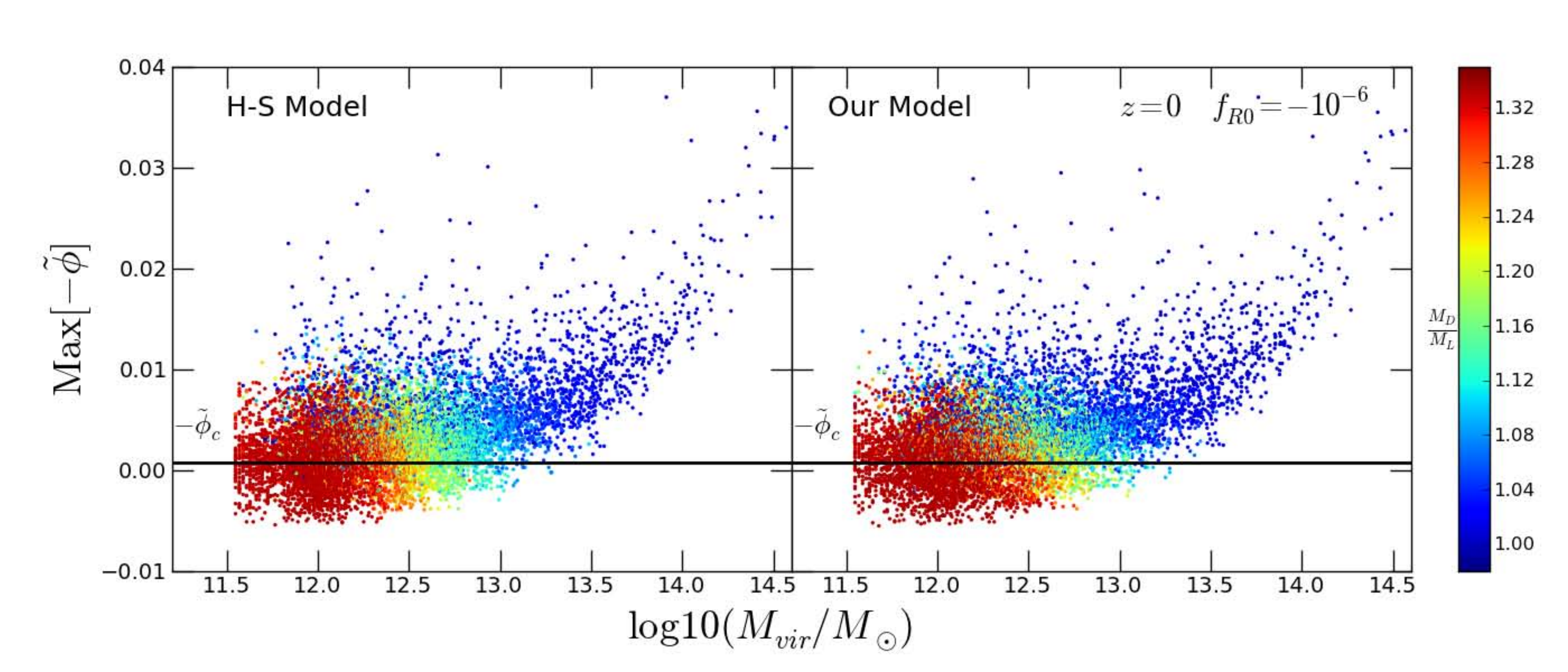}
\caption{Scatter plot for the maximal value of the gravitational potential $\rm Max[-\phi]$ inside a dark halo with respect to the lensing mass of the halo for $f(R)$ models with $f_{R0}=-10^{-6}$ at $z=0$. In this case, most of the massive haloes (e.g. $M_{vir}>10^{13}M_{\odot}$) can generate deep enough potential well and get \tcr{self}-screened. It is also obvious that \tcr{a substantial fraction} of the small haloes are also well screened due to the environment-screening. Below the horizontal line\tcr{, which represents} the critical potential $|\tilde{\phi}_c|=2\tilde{c}^2|\bar{f}_R|$, most of the haloes are completely unscreened.
 }\label{F6halo}
\end{figure*}


We now present our results for several representative models. Show in Figs.~\ref{F4halo}, \ref{F5halo}, \ref{F6halo} are the numerical results for the two $f(R)$ models with $f_{R0}=-10^{-4}$ at $z=1$ (In Fig.~\ref{F4halo}, note that we do not show the $z=0$ results for $f_{R0}=-10^{-4}$, because all haloes in this case are simply unscreened) and the models with $f_{R0}=-10^{-5}$, $f_{R0}=-10^{-6}$ at $z=0$, respectively. In these figures, each point represents a dark matter halo and the colour of the point describes the ratio between the dynamical mass and the lensing mass.
We find the maximal value of the gravitational potential $-\phi$ inside a dark matter halo and show 
$\rm Max[-\phi]$ with respect to the lensing mass of the said halo. For convenience, the potential $\tilde{\phi}$ is in code units, and $\tilde{\phi}_c=2\tilde{c}^2\bar{f}_R$ is the critical potential we have defined in the previous section. 
From 
these figures, we can see that if $-\tilde{\phi}>0$, the completely screened dark matter haloes ($\frac{M_{D}}{M_{L}}\approx 1.0$) only appear in potentials much deeper than 
the critical potential $\tilde{\phi}_c$. It is also evident that below
this critical potential, almost all the haloes are completely unscreened ($\frac{M_{D}}{M_{L}}\approx \frac{4}{3}$). These observations apply to both $f(R)$ models under consideration and for different values of $f_{R0}$. 
\begin{figure*}
\includegraphics[width=6.5in,height=2.7in]{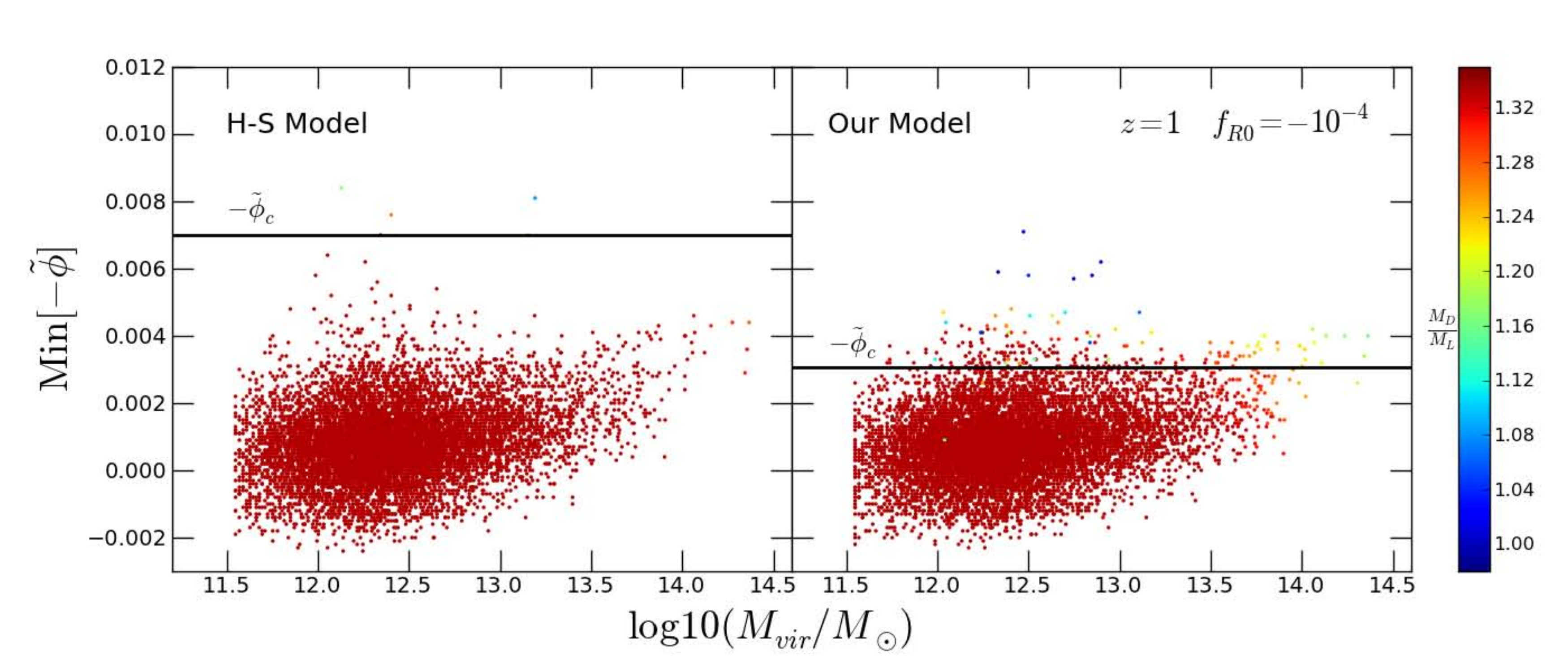}
\caption{Scatter plot for the minimal value of the gravitational potential $\rm Min[-\phi]$ inside a dark halo with respect to the lensing mass of the halo for $f(R)$ models with $f_{R0}=-10^{-4}$ at $z=1$. The small haloes indicated by the blue points \tcr{are embedded in potential wells significantly deeper than} the threshold $|\tilde{\phi}_c|$\tcr{, and} are therefore well-screened. However, the minimal depth of the potential well $\rm Min[-\phi]$ for the massive haloes are not far above the threshold of the potential $|\tilde{\phi}_c|$. \tcr{These} massive haloes are only partially screened (e.g. $\frac{M_{D}}{M_{L}}\sim 1.20$).  }\label{F4haloesmall}
\end{figure*}

Next, we look at the two different ways of screening haloes as mentioned before. The efficiency of the screening depends on the depth of the potential well, $|-\phi|$. In the $f_{R0}=-10^{-4}$ case, as is shown in Fig.~\ref{F4halo}, the dark matter haloes, even the largest ones, cannot generate a deep enough potential well for self-screening, and most of them are completely unscreened. However, we also see that there are several small haloes that are well screened. In these cases, the screened haloes are environmentally screened, because they reside in deep potential wells generated by \tcr{nearby structures}. In order to confirm this point, in Fig.~\ref{F4haloesmall}, we show the minimal values of the gravitational potential $-\phi$ found inside dark matter haloes with respect to the lensing mass of the haloes. \tcr{Compared} with Fig.~\ref{F4halo}, for the large haloes, we find that although the maximal 
depth of the potential well ($\rm Max[-\phi]$) inside the haloes is far above the critical potential, the minimal depth 
$\rm Min[-\phi]$ can be below it: the large haloes are therefore only partially screened, leading to $\frac{M_{D}}{M_{L}}>1$. On the other hand, for the well-screened small haloes, from Fig.~\ref{F4haloesmall}, we can see that even the minimal depths of the potential inside the haloes are far above the critical potential (see the blue points in Fig.~\ref{F4haloesmall}): since the small haloes themselves could not produce such deep potentials, the latter should have been generated by their environments (note that the results are unlikely to be noise as a halo normally contain at least hundreds of simulation particles).

If the background field $|\bar{f}_R|$ is small (e.g., $f_{R0}=-10^{-6}$), most haloes can generate relatively deeper potential wells than the small critical potential $\tilde{\phi}_c$ and thus easily be self-screened. 
From Fig.~\ref{F6halo}, we find that all haloes more massive than about $10^{13}M_\odot$ are well screened. However, not all the small haloes
less massive than $\sim10^{13}M_\odot$ are unscreened. As explained in the above, there are a substantial fraction of the small haloes which are
environmentally screened: as the critical potential $|{\phi}_c|$ is smaller for $f_{R0}=-10^{-6}$, there will be more regions in which nearby \tcr{structures} can create a potential well deeper than $|\phi_c|$.

\begin{figure*}
\includegraphics[width=6.5in,height=2.7in]{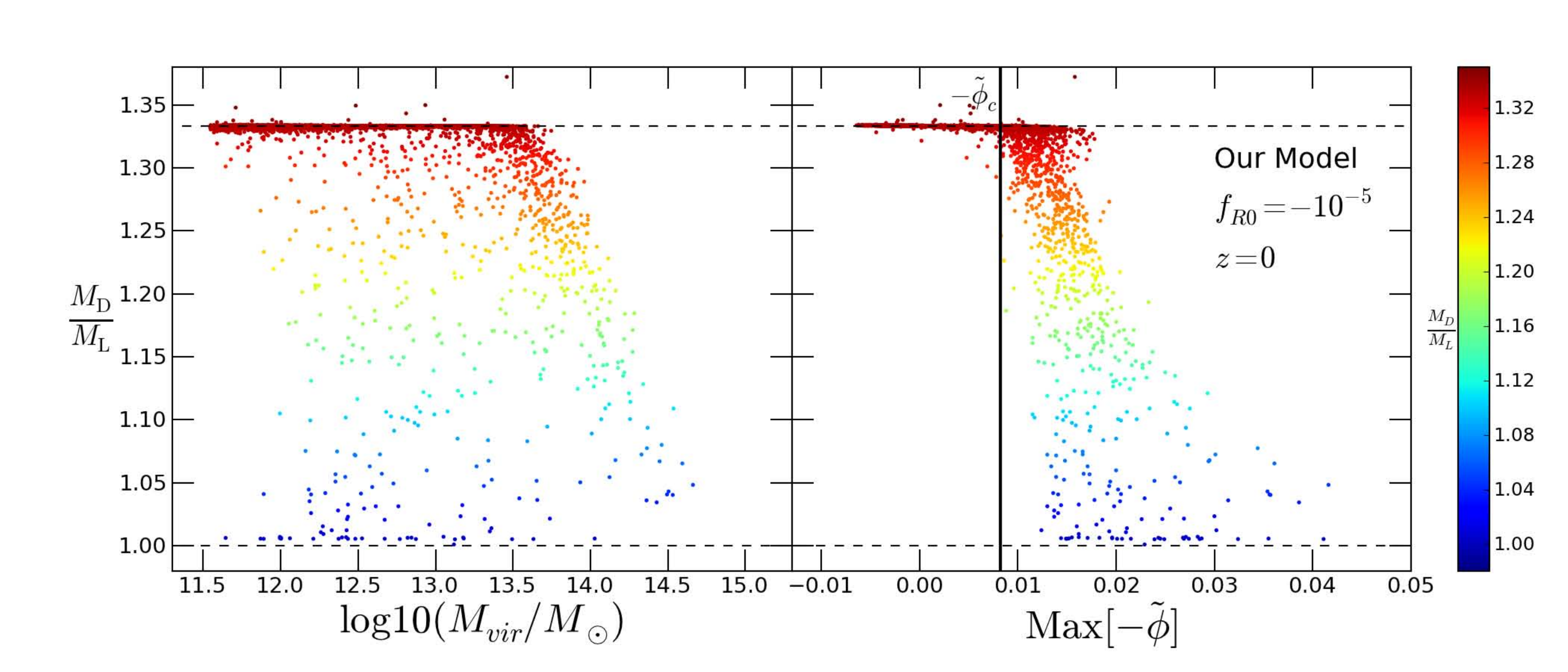}
\includegraphics[width=6.5in,height=2.7in]{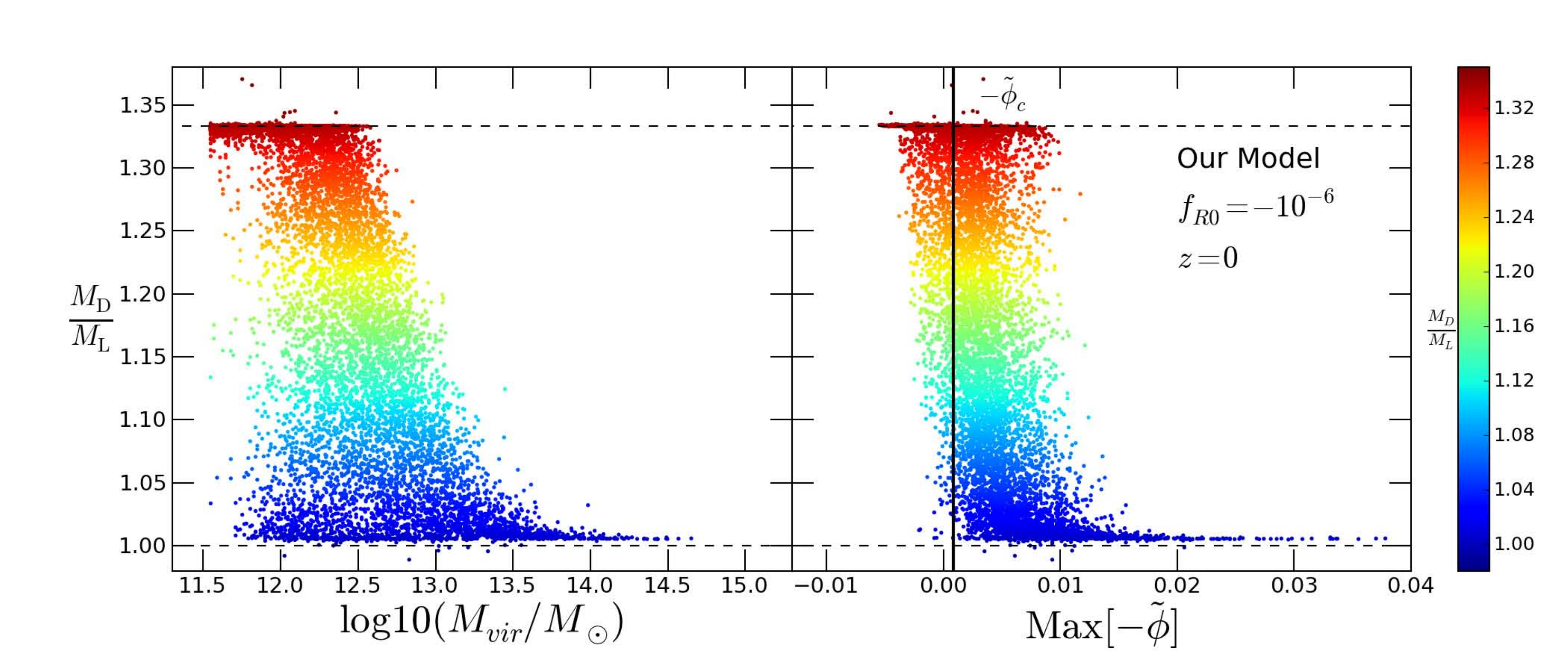}
\includegraphics[width=6.5in,height=2.7in]{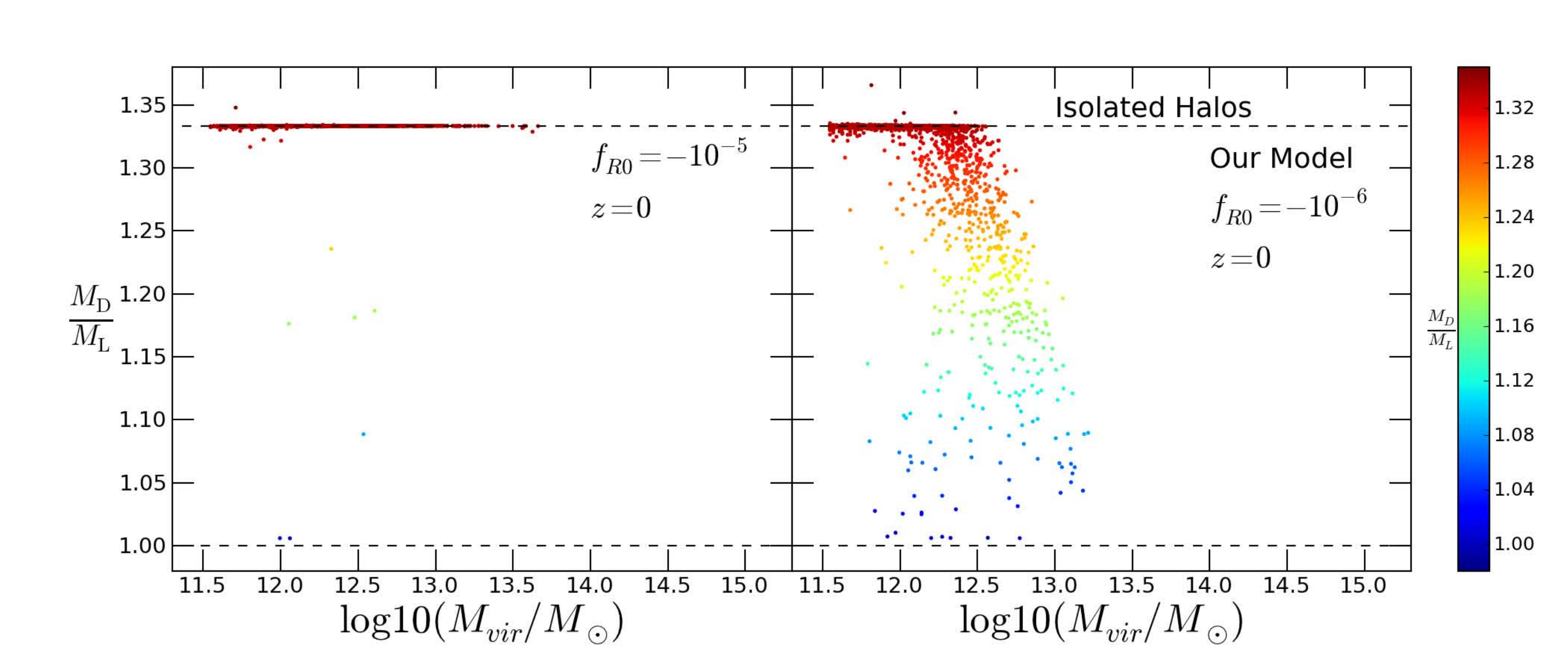}
\caption{Scatter plot for $\frac{M_{\rm D}}{M_{\rm L}}$ with respect to the lensing mass and the maximal value of the gravitational potential $\rm Max[-\phi]$ of each halo for $f(R)$ model at $z=0$ with $f_{R0}=-10^{-5}$ (top row), and $f_{R0}=-10^{-6}$ (middle row), respectively. Statistically speaking, the screening level also depends on the depth of the potential well. It is clear that the scatter in terms of halo mass is much lager than that of the potentials. The transition from unscreened to screened haloes is much sharper in $f_{R0}=-10^{-6}$ case than that in the $f_{R0}=-10^{-5}$ case. The bottom row shows the distribution of $\frac{M_{\rm D}}{M_{\rm L}}$  for isolated haloes. It is evident that most of the isolated haloes are completely unscreened haloes. In $f_{R0}=-10^{-6}$ case, the
screening level shows clear dependence on the halo mass. This is because for isolated haloes the potential is more dependent on its own mass.
  }\label{screeing}
\end{figure*}

Further, statistically speaking, the screening level $\frac{M_{\rm D}}{M_{\rm L}}$ also depends on the depth of the potential well. In Fig.~\ref{screeing} we show the
distribution of $\frac{M_{\rm D}}{M_{\rm L}}$ with respect to $M_{\rm L}$ and the maximal value of the gravitational potential, $\rm Max[-\phi]$, of each halo. It is evident that the scatter of $\frac{M_{\rm D}}{M_{\rm L}}$ as functions of the potentials is much smaller than that of the halo masses. There are clear statistical transition features of haloes from being completely unscreened to being very well screened as the potential deepens. The transition is much sharper in $f_{R0}=-10^{-6}$ than in $f_{R0}=-10^{-5}$, which is as expected given that the condition $|-\phi|\gtrsim|-\phi_c|$ can be more easily satisfied in $f_{R0}=-10^{-6}$ case.

In the bottom row of Fig.~\ref{screeing}, we show the screening level for isolated haloes with respect to lensing masses for $f_{R0}=-10^{-5}$ and $f_{R0}=-10^{-6}$, respectively. The isolated halo is defined as a halo with no neighbours around, by
\begin{equation}
\left|\left|\vec{r}_i-\vec{r}_j\right| \right|\geq N(R_{\rm vir}^i+R_{\rm vir}^j),
\end{equation}
in which $\vec{r}_i$ is the position of a halo's centre and $R_{\rm vir}^i$ is its viral radius; $N$ characterises the separation of haloes, and we take $N = 10$ in this work. From the bottom row of Fig.\ref{screeing}, it is clear that most of the isolated haloes are completely unscreened. In the $f_{R0}=-10^{-6}$ case, the screening level shows clear dependence on the halo mass, because for isolated haloes the screening is mainly self-screening, determined by the halo mass.

\begin{figure*}
\includegraphics[width=6.5in,height=2.7in]{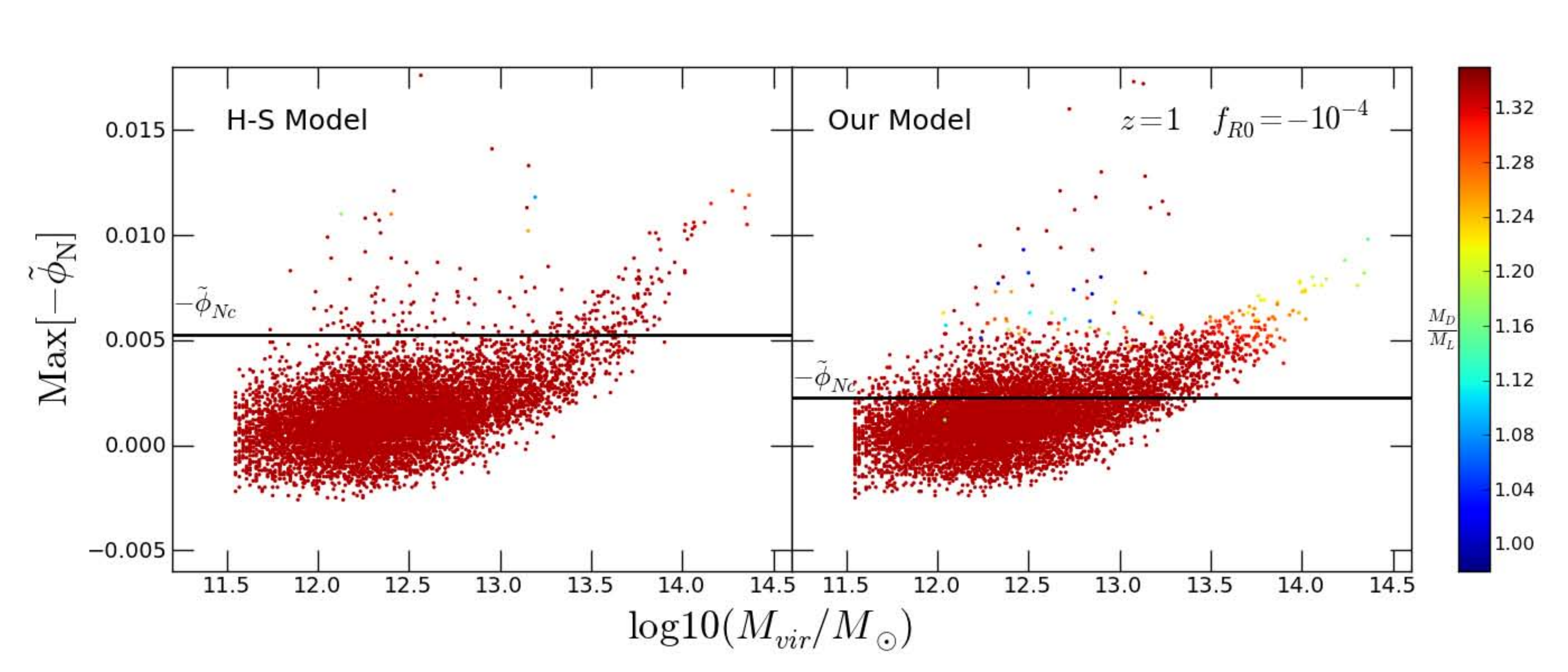}
\includegraphics[width=6.5in,height=2.7in]{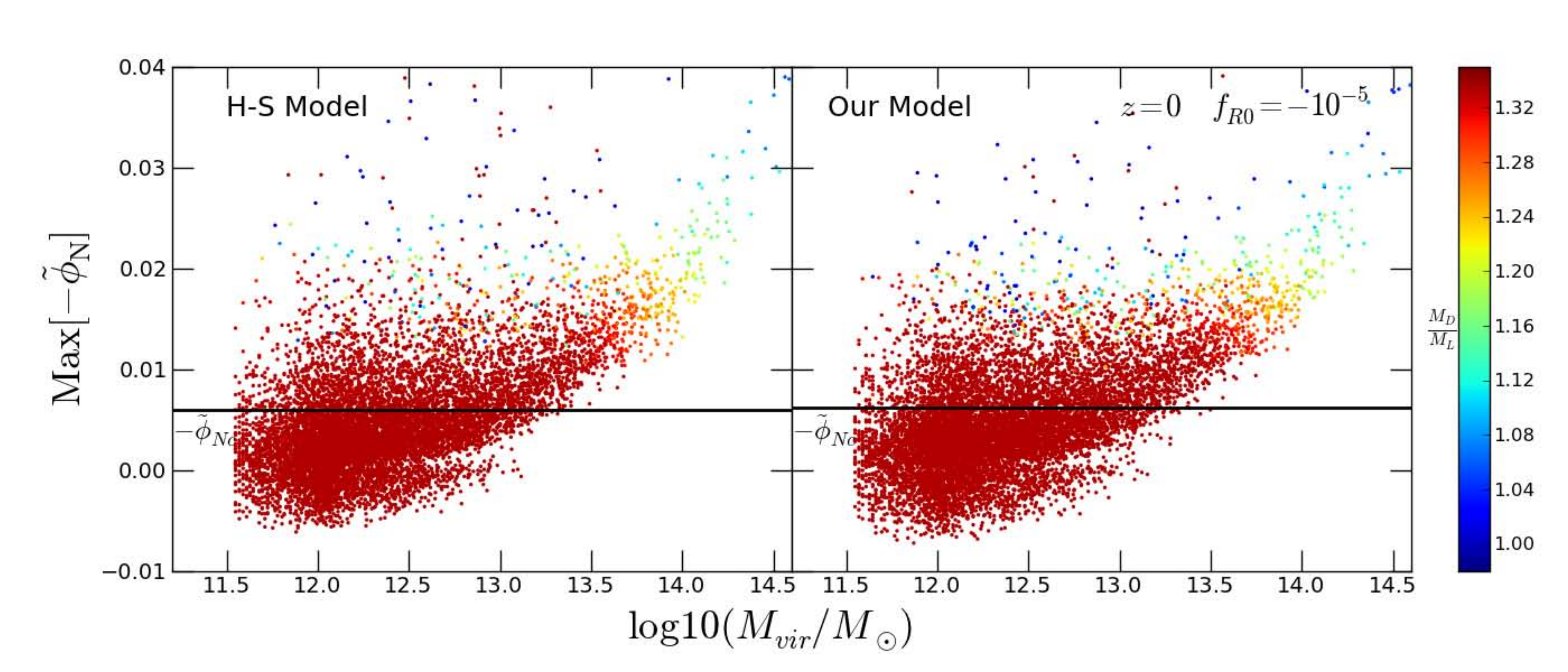}
\includegraphics[width=6.5in,height=2.7in]{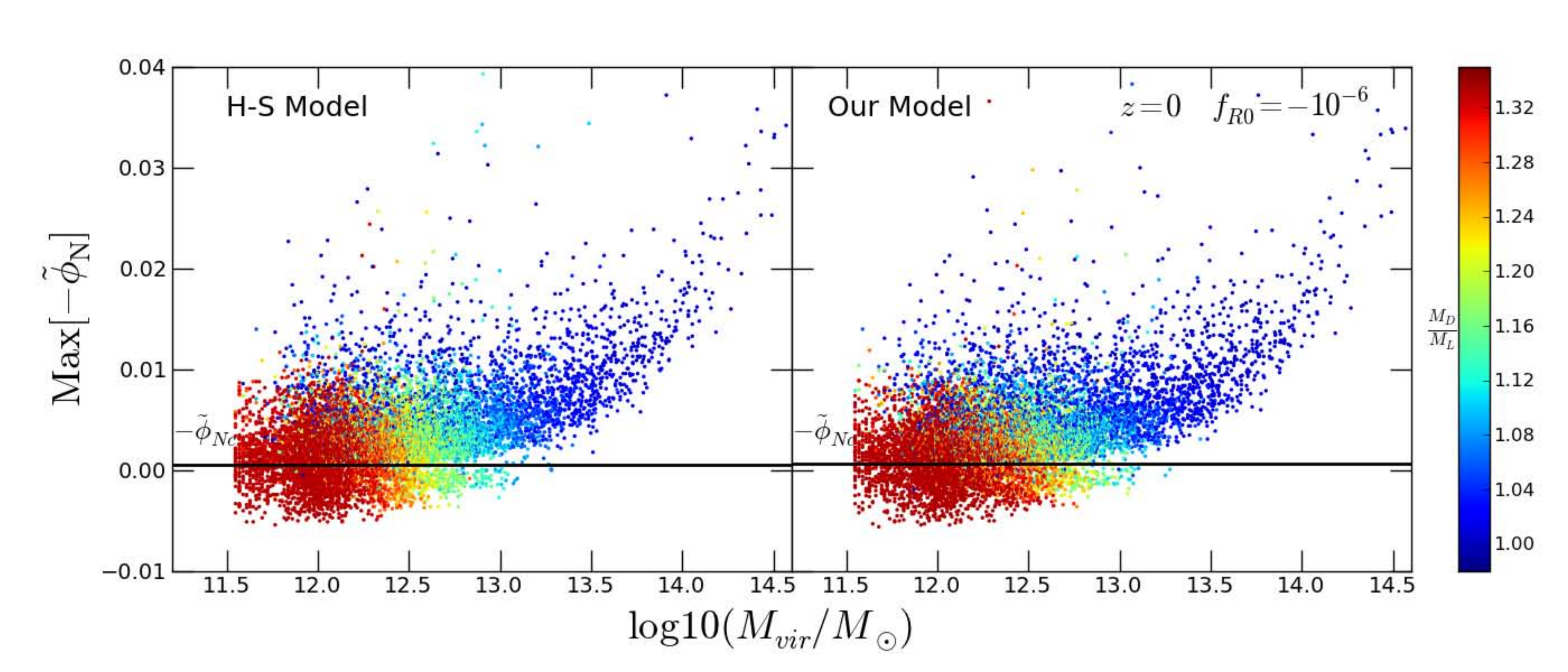}
\caption{Scatter plot for the maximal value of the standard Newtonian potential $\rm Max[-\phi_N]$ inside a dark halo with respect to the lensing mass of the halo for $f(R)$ models with $f_{R0}=-10^{-4}$ at $z=1$ \tcr{(top row)}, and $f_{R0}=-10^{-5}$ \tcr{(middle row)}, $-10^{-6}$ \tcr{(bottom row)} at $z=0$, respectively. The values of the Newtonian potential are evaluated by $\phi_N=\phi+\frac{c^2\delta f_R}{2}$ and $\tilde{\phi}_N$ is in the code units. The critical potential is defined by $\tilde{\phi}_{Nc}=\frac{3}{2}\tilde{c}^2\bar{f}_R$. From the plots, we can see clearly that $|\tilde{\phi}_N|>|\tilde{\phi}_{Nc}|$ is not accurate enough for identifying the screened haloes.
However, $|\tilde{\phi}_N|<|\tilde{\phi}_{Nc}|$ is accurate for identifying the unscreened haloes in the $f_{R0}=-10^{-4}$ and $f_{R0}=-10^{-5}$ cases. In the $f_{R0}=-10^{-6}$ case, below the horizontal line, most of the haloes are completely unscreened, \tcr{though} several \tcr{of them} are only partially unscreened.
  }\label{F6Nhalo}
\end{figure*}

So far, our analysis of the screening mechanism is based on comparing the local gravitational potential $-\phi$, to the value of the background field $c^2\bar{f}_R$. The condition $|-\phi|\lesssim 2c^2|\bar{f}_R|$ is \tcr{useful} for identifying unscreened haloes \tcr{theoretically}. However, \tcr{in practice} a global map of potential $-\phi$ may not be easily constructed in real galaxy surveys, and we need to use the standard Newtonian potential $\phi_N$, namely the lensing potential, which is related to $\phi$ by Eq.~(\ref{Nphi}). There are two reasons for this:
\begin{itemize}
\item First, a global map of $\phi_N$ can be easily constructed in real galaxy surveys if a group catalog \cite{groupcatalog} is available, \tcr{because} $\phi_N$ satisfies the linear equation, Eq.~(\ref{poissonN}). \tcr{$\phi$, on the other hand, can not be reconstructed without solving the more complicated nonlinear scalar field equation.}
\item Second, measurements of galaxy shear also have the potential to reconstruct the 3-dimensional map of the lensing potential $\phi_N$ using weak lensing tomography \cite{WL}.
\end{itemize}

\begin{figure}
\includegraphics[width=3.5in,height=3in]{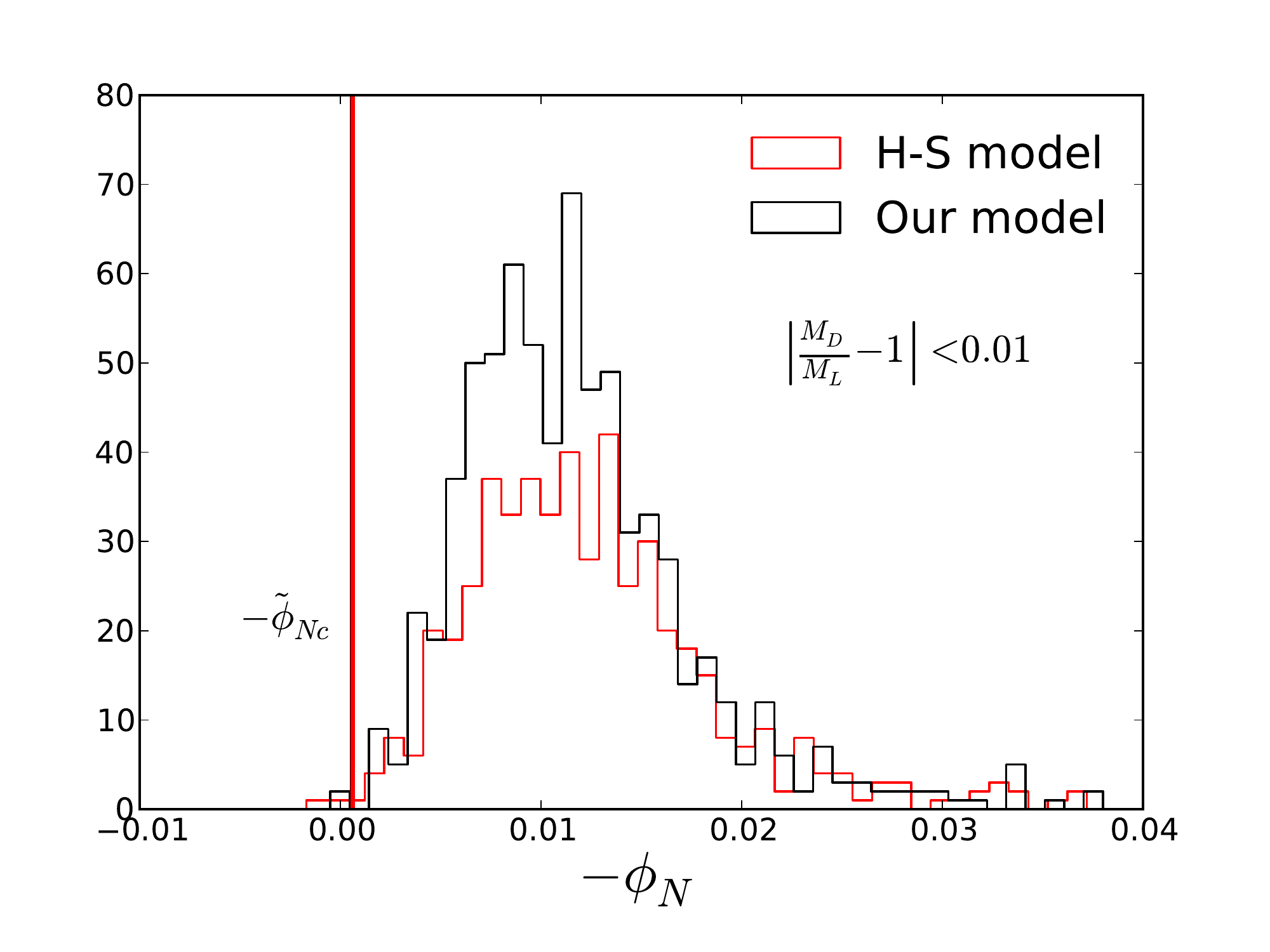}
\caption{Histogram for the well-screened dark haloes ($\left|\frac{M_D}{M_L}-1\right|<0.01$) with respect to the maximal potential $-\phi_N$ inside \tcr{them} for the $f_{R0}=-10^{-6}$ cases. It is clear that almost all the well-screened dark haloes lie above the critical potential $|-\phi_{Nc}|$, \tcr{and b}elow the threshold $|-\phi_{Nc}|$ the number counts are \tcr{very small}.
 }\label{hisgram}
\end{figure}

As we have discussed in the previous section, for identifying the unscreened haloes, the condition $ |-\frac{4}{3}\phi_N|\lesssim2c^2|\bar{f}_R|$ is stronger than $|-\phi|\lesssim 2c^2|\bar{f}_R|$. Let us now examine the power of the condition $ |-\frac{4}{3}\phi_N|\lesssim2c^2|\bar{f}_R|$ for identifying unscreened haloes.
In Fig.~\ref{F6Nhalo}, we show the maximal value of the Newtonian potential $\phi_N$ inside a halo ($\rm Max[-\phi_N]$) with respect to the lensing mass of the halo for $f_{R0}=-10^{-4}$ models at $z=1$ (top panel) and $f_{R0}=-10^{-5}$ (middle panel) and $f_{R0}=-10^{-6}$ (bottom panel) models at $z=0$. The horizontal lines indicate the critical potentials for the Newtonian potential $\phi_N$, which is defined by
\begin{equation}
\phi_{Nc}=\frac{3}{2}c^2\bar{f}_R.
\end{equation}
We can see that $|\phi_N|>|\phi_{Nc}|$ is not 
very useful for identifying screened haloes. However, the opposite case $|\phi_N|<|\phi_{Nc}|$ is very accurate for identifying completely unscreened haloes in $f_{R0}=-10^{-4}$ and $f_{R0}=-10^{-5}$ cases. For $f_{R0}=-10^{-6}$ cases, as shown in Fig.~\ref{F6Nhalo}, not all haloes with 
${\rm Max} [-\phi_{N}]<|-\phi_{Nc}|$ are completely unscreened: several of them (mainly the more massive ones) are only partially unscreened. However, the condition $|\phi_N|<|\phi_{Nc}|$ in this case does distinguish 
unscreened haloes (including partially unscreened ones) from {\it well-screened} haloes 
(dark blue points in Fig.~\ref{F6Nhalo}). In order to show this point, in Fig.~\ref{hisgram} we present a histogram for the distribution of the well-screened dark haloes ($\left|\frac{M_D}{M_L}-1\right|<0.01$) with respect to the maximal potential $-\phi_N$ inside the halo. It is clear that below the threshold $|-\phi_{Nc}|$, the number counts of well-screened haloes are fairly low.

\section{Summary and Discussion\label{con}}

The \tcr{chameleon} screening plays an important role in the \tcr{viability of} $f(R)$ gravity. In this paper, we have reexamined the screening in $f(R)$ cosmology using a suite of $N$-body simulations and found a number of
useful results, which are summarised as follows.
\begin{itemize}
\item In low-density regions, we find that the local curvature $R$
has a nonzero lower bound given by
   \begin{equation}
    R>4\Lambda.
   \end{equation}
   This conclusion applies to a large family of $f(R)$ models that can closely mimic the $\Lambda$CDM background expansion regardless the functional form of $f(R)$. A practical application of this result is that the approximation for the scalar field $f_R$ only needs to be accurate in the range $R>4\Lambda$.

\item In high-density regions, we find an inequality
\begin{equation}
c^2\left|\delta f_R\right|\leq \left|-\frac{\phi}{2}\right|,
\end{equation}
that plays an important role in understanding the screening. We find that screening happens only if the depth of the local potential, $-\phi$, is close to or above the value of the background field, namely $|-\phi|\gtrsim2c^2|\bar{f}_R|$. However, this condition is not sufficient for all haloes to be well screened. On the other hand, we find that the opposite case, $|-\phi|\leq 2c^2|\bar{f}_R|$, can be reliably used to identify completely unscreened haloes in the simulations.

To make connection between our results and real galaxy surveys, we have also expressed the condition in terms of the standard Newtonian potential $\phi_N$, or the lensing potential, which can be more straightforwardly inferred from observations. We show that
\begin{equation}
\left|-\frac{4}{3}\phi_N\right|\leq 2c^2\left|\bar{f}_R\right|,\label{inequality}
\end{equation}
is a stronger and more conservative criterion to find unscreened haloes.
It works very well in the $f_R=-10^{4}$ and $f_R=-10^{-5}$ cases, for which below the threshold
potential $|\phi_{Nc}|=\frac{3}{2}c^2|\bar{f}_R|$ all our dark matter haloes are completely unscreened. In the case of $f_R=-10^{-6}$, although the criterion in Eq.~(\ref{inequality}) no longer guarantees that all the selected haloes are completely unscreened, it does cleanly separates unscreened haloes from the well-screened ones, and the contamination of the unscreened samples is very low.
\end{itemize}

We point out that the way we separate self and environmental screenings of dark matter haloes is slightly different from some works in the literature. When talking about environmental screening, people often use a criterion similar to Eq.~(\ref{inequality}), but with (i) $\bar{f}_{R}$ replaced by its `local' version $\bar{f}_{R,\xi}$, where the subscript $\xi$ means that $\bar{f}_{R,\xi}$ is the average over a region of size $\xi$, usually assumed as comparable to the Compton wavelength of $f_R$, and (ii) $\phi_N$ taken as the Newtonian potential {\it generated by} the object (halo or galaxy) being considered, instead of the total Newtonian potential {\it measured near} the said object (the latter could have contribution from nearby objects). Our criterion is more directly related to observations, as we can only measure the total $\phi_N$ with gravitational lensing -- if the latter is known, we know the total screening and the separation of self and environmental screening is of no practical interest. Furthermore, note that $\phi_N$ satisfies the usual superposition principle, while $\phi$ does not, and thus the use of $\phi_N$ makes it easier to estimate the contributions from environment (neighbouring structures).

Although our conclusions are based on pure dark matter simulations, we would like to point out that the screening of a galaxy should be generally determined by the screening of the underling dark matter field since the baryon field only accounts for a small fraction of the total matter field on the scale of halos.

State-of-the-art hydro simulations in the standard $\Lambda$CDM model, such as the {\sc Eagle} \cite{eagle} project, have led to clear pictures of the baryon distribution in dark matter haloes. For illustrative purposes, we assume that this picture also roughly holds for $f(R)$ gravity. The baryon contribution to halo masses is just $\sim2-3\%$ for haloes of $\sim10^{11}M_{\odot}$, rising gradually to $\sim15$\% for haloes of $\sim10^{14}M_\odot$. It is clear that the baryons only account for a small fraction of the halos mass and further only a small fraction of baryons come into the form of the stellar mass.

Within dark matter haloes, dark matter dominates the matter field when the radius is above $5\%$ halos radius $r>0.05R_{\rm vir}$, where the density profile is well described by the NFW profile. In the core part of the halo $r<0.05R_{\rm vir}$, baryons would make up a significantly larger fraction of the total masses, at $\sim10-25\%$ for haloes of $10^{11}-10^{14}M_\odot$. In this region, baryons are almost completely in the form of stars. Although the total density profile in this region is deeper than NFW, using the fitting results of \cite{eagle}, we find that the baryons still contribute a sub-dominate fraction to a halo's own potential $\phi_N$. It is about $\sim25\%$ in haloes of $10^{11}M_\odot$, rising to $\sim40\%$ for haloes of $\sim10^{12}M_\odot$ and then decrease to below $\sim10\%$ for haloes of $10^{14}M_\odot$. The presence of a galaxy near the halo centre therefore will not dramatically change the screening property therein, though it can make a quantitative difference.

Although the screening properties on the scale of a galaxy is determined by the dark matter field, it is important to note that on the scale of stars, the screening is determined by the baryon field itself since dark matters can not be localized in such a small dense region. If the star is dense enough, the potential in the center region will be very deep. The star, at least, will be partially self-screened. It is very interesting to note that, in an unscreened halo, the stars can be treated as if living on the cosmological background. We take $f_{R0}=-10^{-6}$ for instance. If a halo is unscreened, it means that its $|\phi_N|$ is smaller than $\sim10^{-6}$, the halo has a mass of $\sim10^{12}M_\odot$ and its baryons contribute an additional potential of $\phi_N\sim-5\times10^{-7}$, which is still not enough to screen the halo. The contribution from the halo and galaxy to the potential of stars can be neglected, which, in turn, means that the screening of a star is determined by the depth of its own potential relative to the cosmological background field $|\bar{f}_{R0}|$.

Comparing the properties of galaxies in screened versus unscreened haloes could potentially provide one of the most robust tests of $f(R)$ gravity \cite{Jain,cabre,hui}, because the formation and evolution of galaxies in these regions should differ significantly due to the $1/3$ enhancement of the gravitational force. However, caution must be taken when performing and interpreting these tests, due to the difficulty of correctly modelling the nonlinear environmental effects. Detailed simulations and analysis of galaxy formation in $f(R)$ gravity are needed before drawing quantitative conclusions.

When making applications to real galaxy surveys, the first step is to build a screening map \cite{cabre}. The unscreened samples are of particular interest. 
As is discussed above, massive components in the galaxy, such as stars, can self screen if they can generate deep enough local potential wells such that $|\phi_N|\gg|\phi_{Nc}|$, where the threshold potential $|\phi_{Nc}/c^2|$ for models with different values of $f_{R0}$ at $z= 0$ are listed in Table~\ref{values}. Here remember that $\phi_N$ has additional contributions from the galaxy, its host halo and their large-scale environment. The Sun typically has the potential as $|\phi_{N\odot}/c^2|\sim10^{-6}$ and consequently main sequence stars similar to or more massive than the Sun could be at least partially self-screened for $f(R)$ models with $|f_{R0}|\leq10^{-6}$. Only low density components like the gaseous disk and low-mass stars, in unscreened haloes, are unscreened. This picture of partially-screened galaxy opens a novel opportunity to test $f(R)$ gravity by examining the different dynamics between their screened and unscreened components~\cite{Dwarf}.

However, as is pointed out in this work, to accurately identify unscreened galaxies in real surveys, we need to estimate the {\it total} Newtonian potential $\phi_{N}$ at the positions of the galaxies, considering the latter to be tracers of the underlying dark matter field. A group catalogue could be used for this kind of study (e.g., Ref.~\cite{cabre}), and it is crucial to understand how well the group
luminosity of galaxy samples can trace the underling dark matter halo mass in $f(R)$ gravity. When converting the group luminosity to the halo mass, further caution must be taken because there may be significant difference in the biases of screened and unscreened haloes. This work requires a careful investigation of halo and galaxy formation in $f(R)$ gravity and therefore higher resolution simulations, which will be addressed in our future work.

\begin{table}
\caption{$|\phi_{Nc}/c^2|$ for $f(R)$ models}\label{values}
\begin{tabular}{c|c}
  \hline
  \hline
  $f_{R0}$ & $\phi_{Nc}/c^2=\tilde{\phi_{Nc}}/\tilde{c}^2=\frac{3}{2}f_{R0}$ \\
 \hline
  $-10^{-4}$ & $-1.5\times10^{-4}$\\
  $-10^{-5}$ & $-1.5\times10^{-5}$\\
  $-10^{-6}$ & $-1.5\times10^{-6}$ \\
  \hline
\end{tabular}
\end{table}

Furthermore, the galaxy shear measurements may have the potential of determining the Newtonian potential $\phi_N$, namely the lensing potential, with significantly improved precisions. Coming surveys such as Euclid~\cite{Euclid} will be able to reconstruct the three-dimensional lensing potential using weak lensing tomography \cite{WL}. With these, the method presented in this paper offers a reliable way to select unscreened samples from galaxy surveys. Combining galaxy shear measurements, galaxy surveys and additional observations on the galaxy properties may yield powerful tests on $f(R)$ gravity in the future.

Finally, we would like to remark here that the efficiency of screening depends on the absolute depth of the potential well. This is due to the non-linear nature of the scalar field equation Eq.~(\ref{frpoisson}). The reference of the depth of the potential well $\delta f_R$ can not be chosen arbitrarily because $\delta f_R$ should vanish for the homogenous density field, which actually defines the zero point of $\delta f_R$. The Newtonian potential, $\phi_{N}$, should vanish for the homogenous density field as well. To apply our results to real galaxies surveys, we need to carefully take into account this point.

\begin{acknowledgments}

We thank L.~Guzzo for helpful discussions. JHH acknowledges support of the Italian Space Agency (ASI), through contract agreement I/023/12/0. BL is supported by the Royal Astronomical Society and Durham University.
AJH and BRG acknowledge support of the European Research
Council through the Darklight ERC Advanced Research Grant (291521).

\end{acknowledgments}

\end{document}